\documentclass[12pt,preprint]{aastex}
\newcommand{\doctype}{paper}

\newcommand{\bFW}{\textit{F606W}}

\newcommand{\bR}{\textit{R}}

\newcommand{\bJ}{\textit{J}}
\newcommand{\bH}{\textit{H}}
\newcommand{\bK}{\textit{K}}
\newcommand{\bKp}{\textit{K}$'$}
\newcommand{\bL}{\textit{L}}
\newcommand{\bLp}{\textit{L}$'$}

\newcommand{\AUMic}{AU~Mic}
\newcommand{\betaPic}{$\beta$~Pic}

\newcommand{\MCFOST}{\textsc{MCFOST}}
\newcommand{\NextGen}{\textsc{NextGen}}

\newcommand{\x}{\ensuremath{\mathbf{x}}}
\newcommand{\tausca}{\ensuremath{\tau_\perp^\mathrm{sca}}}
\newcommand{\taugeo}{\ensuremath{\tau_\perp^\mathrm{geo}}}

\defcitealias{liu04}{L04}
\defcitealias{krist_etal05}{K05}
\defcitealias{metchev_etal05}{M05}
\defcitealias{strubbe&chiang06}{SC06}

\begin{document}
\slugcomment{\textsc{ApJ accepted:} 2007 May 27}

\title{The AU\,Microscopii Debris Disk: Multiwavelength Imaging and Modeling}
\author{Michael P. Fitzgerald\altaffilmark{1,2,3},
Paul G. Kalas\altaffilmark{1,3},
Gaspard Duch{\^e}ne\altaffilmark{4},
Christophe Pinte\altaffilmark{4}, and
James R. Graham\altaffilmark{1,3}}
\altaffiltext{1}{Department of Astronomy, 601 Campbell Hall, University of California, Berkeley, CA 94720} 
\altaffiltext{2}{\texttt{fitz@astro.berkeley.edu}}
\altaffiltext{3}{National Science Foundation Center for Adaptive Optics, University of California, Santa Cruz, CA 95064}
\altaffiltext{4}{Laboratoire d'Astrophysique, Observatoire de Grenoble, BP 53, F-38041 Grenoble Cedex 9, France}

\begin{abstract}
Debris disks around main sequence stars are produced by the erosion and evaporation of unseen parent bodies, which are potential building blocks of planets.  Such planets may imprint dynamical signatures on the structure of the debris disk.  AU\,Microscopii (GJ\,803) is a compelling object to study in the context of disk evolution across different spectral types, as it is an M dwarf whose near edge-on disk may be directly compared to that of its A5V sibling \betaPic.  We resolve the disk from 8--60\,AU in the near-IR \bJ\bH\bKp\ bands at high resolution with the Keck II telescope and adaptive optics, and develop a novel data reduction technique for the removal of the stellar point spread function.  The point source detection sensitivity in the disk midplane is more than a magnitude less sensitive than regions away from the disk for some radii.  We measure a blue color across the near-IR bands, and confirm the presence of substructure in the inner disk.  Some of the structural features exhibit wavelength-dependent positions.

The disk architecture and characteristics of grain composition are inferred through modeling.  Previous efforts have modeled the dust distribution through a variety of means, ranging from power law models to calculations of steady-state grain dynamics.  Recent measurements of the polarization properties of the scattered light indicate the presence of porous grains.  The scattering properties of these porous grains have a strong effect on the inferred structure of the disk relative to the majority of previously modeled grain types.  We approach the modeling of the dust distribution in a manner that complements previous work.  Using a Monte Carlo radiative transfer code, we compare a relatively simple model of the distribution of porous grains to a broad data set, simultaneously fitting to midplane surface brightness profiles and the spectral energy distribution.  Our model confirms that the large-scale architecture of the disk is consistent with detailed models of steady-state grain dynamics.  Here, a belt of parent bodies from 35--40\,AU is responsible for producing dust that is then swept outward by the stellar wind and radiation pressures.  We infer the presence of very small grains in the outer region, down to sizes of $\sim$0.05\,\micron.  These sizes are consistent with stellar mass-loss rates $\dot{M}_*\ll 10^2\,\dot{M}_\sun$.
\end{abstract}


\section{INTRODUCTION}\label{sec:intro}

In the past two decades, observations enabled by new technology have substantially broadened our understanding of dust surrounding main-sequence stars.  These tenuous disks of dust, found around nearby stars, are fundamentally linked to the processes of star and planet formation~\citep{backman&paresce93, lagrange_etal00, zuckerman01, meyer_etal07}.  Current scenarios favor casting ``debris disks'' in the later stages of the systems' formation, after most of the primordial gas and dust have dissipated~\citep[on timescales of 1--10\,Myr; e.g.][]{zuckerman_etal95, haisch_etal01}.  The debris is freshly nourished by the sublimation, evaporation, and collisional destruction of orbiting parent bodies~\citep[for a recent review, cf.][]{meyer_etal07}.  Since a portion of these parent bodies may take part in the accretion of cores into planets --- which may in turn gravitationally perturb the dust --- circumstellar debris disks offer evidence for the presence of planets and insight into their formation.

Debris disks are usually discovered by sensing the thermal re-radiation of absorbed starlight in the far IR~\citep{backman&paresce93, decin_etal03, bryden_etal06}.  The spectral energy distributions (SEDs) of these systems, particularly the range from the IR to the sub-mm, are commonly used to infer the (often cold) dust temperature and extent of disk inner clearing~\citep[e.g.][]{chen_etal05}.  However, in the absence of additional information, significant degeneracies exist in modeling the dust disks with SEDs~\citep[and references therein]{moro-martin_etal05}.  Severe ambiguities between disk morphology and the grains' physical characteristics are removed with spatially resolved images of the thermal and scattered light; however, the faintness of the dust and often-large dynamic range between the star and disk present significant obstacles to their acquisition.  Resolved images are both rare and valuable.

Images at multiple wavelengths are yet more powerful.  Spatially resolving the debris across several wavebands provides definitive evidence for the size, space, and compositional distributions of the grains~\citep[e.g.][]{pantin_etal97, wyatt06, golimowski_etal06}.  Coupling such data with detailed models of grain dynamics can expose the fundamental physical phenomena governing the production and diffusion of dust throughout the disk.  Physical models trace the characteristics of grains (i.e. composition and geometry) to the underlying parent body distribution, which in turn enables inferences regarding the coagulation history of solid material.
These parent bodies may be corralled into a specific architecture through the gravity of nearby planets.  Moreover, planets can imprint their signatures on the dust disk itself~\citep[e.g.][]{liou&zook99, wyatt_etal99, ozernoy_etal00, kuchner&holman03, kenyon&bromley04, wyatt06}.
Tracing the mechanisms of grain dynamics through resolved imaging also addresses crucial questions about the nature of dust production in these systems --- is the dust distribution steady-state, or is the evolution of debris governed by transient events, such as rare collisions between massive bodies or extrasolar analogs to the Late Heavy Bombardment?
Again, the images that are critical to addressing these issues are rare, though the number of resolved systems has increased rapidly in the past few years (for recent tabulations, see~\citealp{kalas_etal06} and~\citealp{meyer_etal07}).  With ever-increasing sample sizes, we have entered an era where we can compare the structure and evolution of circumstellar dust with quantities which shape the paths of planet formation, such as stellar mass and metallicity.

Contemporary work has revealed an exciting laboratory for the detailed study of circumstellar debris.  \objectname[GJ 803]{AU\,Microscopii} (GJ\,803) is a nearby star that harbors an optically thin debris disk.  The spectral type in the literature ranges from dM0--2.5Ve~\citep{joy&abt74, linsky_etal82, keenan83, gliese&jahreiss95}.  It is a member of the \objectname[beta Pic]{$\beta$\,Pictoris} moving group, and as such is one of the youngest~\citep[$12^{+8}_{-4}$\,Myr;][]{barradoynavascues99, zuckerman_etal01} and among the closest~\citep[9.94\,$\pm$\,0.13\,pc;][]{perryman_etal97} of the known resolved debris disks.  It is especially attractive to study in the larger context of disk evolution as a function of stellar mass.  In terms of mass, M dwarfs like \AUMic\ constitute the dominant stellar component of the Galaxy.  Despite the abundance of such stars, \AUMic\ is currently unique among the resolved debris systems.  It is a touchstone for studying the evolution of circumstellar disks around low-mass stars.
\AUMic\ has a well-chronicled history of flare activity~\citep[e.g.][]{robinson_etal01}.  It is likely that strong stellar activity has a significant effect on dust dynamics and lifetimes~\citep{plavchan_etal05}.  Notably, we can make direct comparisons across spectral types given \AUMic\ and its sibling \betaPic, the archetypal A star with an edge-on debris disk.

The scattered light of the dust  around \AUMic\ was discovered by~\citet{kalas_etal04}, who used seeing-limited \bR-band coronagraphic imaging to resolve it into an extended, near-edge-on disk.  Additional high-resolution studies followed, with adaptive optics (AO) observations at \bH-band by~\citet[hereafter~\citetalias{liu04}]{liu04} and~\citet[hereafter~\citetalias{metchev_etal05}]{metchev_etal05}, as well as in the visible with the \textit{Hubble Space Telescope} (\textit{HST}) by~\citet[hereafter~\citetalias{krist_etal05}]{krist_etal05}.
These detailed images reveal a very thin midplane (FWHM\,$\sim$\,2\,AU) with an inner disk closely aligned with the line of sight~\citepalias[$\theta\sim 0.5\degr$ for $r< 50$\,AU;][]{krist_etal05}.
The midplane surface brightness profiles show a break around 35--45\,AU, with brightness decreasing more sharply at larger projected distances.  There are slight asymmetries in the overall brightness of the profiles between the two disk ansae.
Further, the midplane exhibits substructure on smaller scales, including localized enhancements and deficits in brightness, and vertical deviations of midplane positions away from that of a uniform disk.
Comparison of these structures between datasets is required to confidently reject image processing artifacts.  The origin of the small-scale structure remains unexplained.
A striking feature of the scattered light disk is its color.
\citetalias{krist_etal05} found a blue color in the visible with \textit{HST}, along with an apparent color gradient --- the disk is increasingly blue from 30--60\,AU.
The disk is also blue from the visible to the \bH\ band, as noted by~\citetalias{metchev_etal05}.
This is unlike many of the disks resolved to date, which are neutral or red scatterers like \betaPic~\citep{golimowski_etal06, meyer_etal07}.  We note, however, that two other recently imaged disks, HD\,32297~\citep{kalas05} and HD\,15115~\citep{kalas_etal07}, also scatter blue between the optical and near-IR.
Such data are important because images at each new wavelength serve to further constrain the sizes, composition, and structure of grains through analysis of their scattering properties.

While the scattered light imaging of the disk has attracted attention in the past few years, there has also been significant study of the grains' thermal emission in the SED.  The cold dust around \AUMic\ was first identified by the \textit{Infrared Astronomical Satellite} (\textit{IRAS}), as excess emission at 60\,\micron\ was weakly detected~\citep{mathioudakis&doyle91, song_etal02}.  Recently, sensitive measurements of the 850\,\micron\ flux confirmed the presence of cold dust around the star~\citep{liu_etal04}.  The \textit{Spitzer Space Telescope} has further constrained the thermal emission of the dust in the mid-IR~\citep{chen_etal05}.

Results of these studies have been taken as indirect evidence for the presence of planets around the star.
The gravitational influence of such planets provides a possible mechanism for generating the substructure in the disk.  In particular, mean-motion resonances from a planet can trap dust to produce radial, azimuthal, and vertical structure~\citep[e.g.][]{ozernoy_etal00,wahhaj_etal03,thommes&lissauer03,wyatt06}, though to date, no models of these mechanisms have been applied to the substructure of \AUMic.
Further, the shallow surface brightness profiles in the inner disk and lack of thermal excess in the 10--20\,\micron\ region of the SED have suggested an inner clearing of dust, perhaps maintained again by the presence of an inner planet~\citep[e.g.][]{roques_etal94, moro-martin&malhotra05}.
Nonetheless, it is imperative to understand the detailed physics affecting the distribution of dust grains around this disk prior to establishing the presence of unseen planets.  The gravity of such perturbers is just one of many effects which may shape the dust distribution, including grain-grain collisions, forces from both stellar radiation and wind, and gas drag.
Because the disk is along the line-of-sight to the star, \AUMic\ is a favorable target for using absorption spectroscopy to search for remnant gas in the disk.   \citet{roberge_etal05} placed limits on the column of H$_2$ toward the star using the \textit{Far Ultraviolet Spectroscopic Explorer} (\textit{FUSE}), concluding that giant planets were unlikely to have formed given the rapid dissipation of primordial gas.
Different groups have placed planet detection limits around the star in very narrow~\citep[$r<2$\arcsec\,$\approx$\,20\,AU;][]{masciadri_etal05} and slightly wider fields~(\citealt{neuhauser_etal03};~\citetalias{metchev_etal05}), and to date none have been directly detected.  Certainly the question of planets in this system warrants further study, as these different lines of evidence have not been resolved.

Comprehending the physics that sculpt the dust distribution in this debris disk remains a key step for not only determining the existence of planets, but also for informing us about the evolution of solid material around stars in general.  \citet{augereau&beust06} and~\citet[hereafter~\citetalias{strubbe&chiang06}]{strubbe&chiang06} have both investigated the observed break in the midplane surface brightness profile and color gradient in the outer disk, and hypothesize that a ring of parent bodies (near the break in surface brightness profile) acts as a source of dust grains, which are subsequently swept outward by a strong stellar wind.  \citetalias{strubbe&chiang06} provide a detailed physical model for \AUMic\ for this scenario.  The steady-state spatial and size distributions of dust grains are determined by a small set of physical processes (e.g. collisions, radiation, wind, etc.), and with other assumptions about grain properties, the model can reproduce observations of scattered light profiles and the SED.

As illustrated by the work done to date, we have constraints on the distribution of dust using scattered light images coupled with the SED.  However, in this edge-on system, a degeneracy remains between the scattering properties of grains and their supposed spatial distribution.  The advantage of measurements in polarized light, rather than total intensity, is that they give complementary information in the grain scattering properties  (for grain sizes $a$ such that $x\simeq 2\pi a/\lambda\sim 1$) and thus reduce this degeneracy.  The polarization properties of \AUMic's disk have recently been measured with \textit{HST}/ACS by~\citet{graham_etal07}.  The disk exhibits strong gradients in linear polarization, and their modeling of the flux of polarized visible light indicates the presence of small porous grains and an architecture consistent with a significant ($>$300:1 in vertical optical depth to scattering) inner clearing.

In this \doctype, we report and analyze multiband AO observations of the scattered light disk of \AUMic\ in the near IR.  Our high-contrast images were processed with a novel data reduction technique which aims to mitigate the effects of point-spread function variability on ground-based observations.  We report on the observed colors and color gradients of the disk brightness profiles for our images, and also for reprocessed \textit{HST} data previously presented by~\citetalias{krist_etal05}.  Guided by the recent characterization of grain scattering properties by~\citet{graham_etal07}, we simultaneously fit a dust model to near IR and visible scattered light data as well as the SED.  With this model of dust distribution, we then check the consistency of~\citetalias{strubbe&chiang06}'s physical model for the disk architecture with our empirical results.  We examine the question of whether the inner disk is populated, as expected if corpuscular drag forces can draw grains inward before they are pulverized through collisions.
We document our observational and data reduction techniques in~\S\ref{sec:obs&reduc}; we present our observational results in~\S\ref{sec:results} and analyze dust models in~\S\ref{sec:analysis}.  Discussion and conclusions follow in~\S\ref{sec:discn&concl}.

\section{OBSERVATIONS \& REDUCTION}\label{sec:obs&reduc}

The achievable spatial resolution is a prime motivator for imaging debris disks in the optical and near infrared --- resolved images here highlight structural details and dust properties.  However, at these wavelengths the star overwhelms the light scattered by the circumstellar debris.  This contrast ratio characterizes a fundamental observational challenge in such studies of debris systems.  To meet this challenge, we use AO to concentrate the star's light and employ a coronagraph to occult the resulting stellar image, thereby increasing sensitivity to the disk.  Here, we detail our use of AO coronagraphy and image processing techniques to study the circumstellar dust~(\S\ref{subsec:nir_im}--\ref{subsec:psf_sub}).  We describe calibration of the AO data in~\S\ref{subsec:calibration}, and our use of \textit{HST} images in~\S\ref{subsec:hst_data}.

\subsection{Near Infrared Imaging}\label{subsec:nir_im}

We observed \AUMic\ on the nights of 2004 Aug 29-30 with the Keck II AO system and a coronagraphic imaging mode of the NIRC2 camera.  \AUMic\ is sufficiently bright to serve as its own reference for adaptive wavefront correction.  After wavefront compensation, the on-axis starlight is blocked by a focal-plane mask.  Diffraction effects are then suppressed by a pupil-plane Lyot stop, and the light is reimaged onto a 1024\,$\times$\,1024\,pixel Aladdin detector.  The narrow-field mode of the camera was used, at a scale of $\approx$\,10\,mas~pixel$^{-1}$.  Exposure times were 30--60\,s in \bJ\bH\bKp\ (60\,$\times$\,1\,s in \bLp), and the filter was cycled after a few short exposures in each band.  Exposures where the disk was aligned with a diffraction spike were discarded.  The filters, total integration times, and focal masks are listed in Table~\ref{tab:obs}.  We used a 0\farcs 5 radius focal plane mask on the second night, however the data in the region around the edge of this mask were discarded due to large residual subtraction errors.
Calibration employed standard bias subtraction and flat fielding techniques.

Imaging faint circumstellar material requires suppression of the stellar light, whose distribution is given by the on-axis point spread function (PSF).  While the coronagraph suppresses much of the starlight, a fraction leaks through the system and must be subtracted from the data in post-processing.  The relative success of this operation is linked to PSF stability.  In observations through an altitude-azimuth telescope like Keck, the image plane rotates relative to the telescope with parallactic angle.  However, features in the PSF that are produced by the telescope (e.g. diffraction spikes) maintain a fixed orientation relative to it.  We disabled the instrument rotator of NIRC2 to additionally fix the orientation of any features arising from camera aberrations.  With the disabled rotator, the PSF orientation is fixed relative to the detector while the edge-on disk appears to rotate around the stellar image with time.  This allows us to disentangle the image of the disk from features in the diffraction pattern of the star, as well as use the stable orientation of the stellar PSF on the detector for more accurate subtraction.  ``Roll subtraction'' is a general observational methodology that has been applied with success to AO imaging of circumstellar disks~\citepalias[e.g.][]{liu04, metchev_etal05}.  An account of our version of the roll subtraction technique follows.

\subsection{PSF Subtraction Technique}\label{subsec:psf_sub}

The time variability of the AO PSF is often the limiting factor in detecting faint circumstellar material from the ground.  Rather than subtracting the stellar PSF from non-contemporaneous observations of reference stars, we wish to directly exploit information from the science target exposures.  Techniques developed by~\citet{veran_etal97} and~\citet{sheehy_etal06} seek to model the AO system to estimate the PSF, using telemetry data and crowded field imaging, respectively.  However, these techniques do not yet have demonstrated applicability to high-contrast imaging.  For roll subtraction, mitigation of AO PSF variability can potentially improve contrast.  We note in particular the technique's advancement in this regard by~\citet{marois_etal06} for the case of point source detection.

\begin{figure}
\plotone{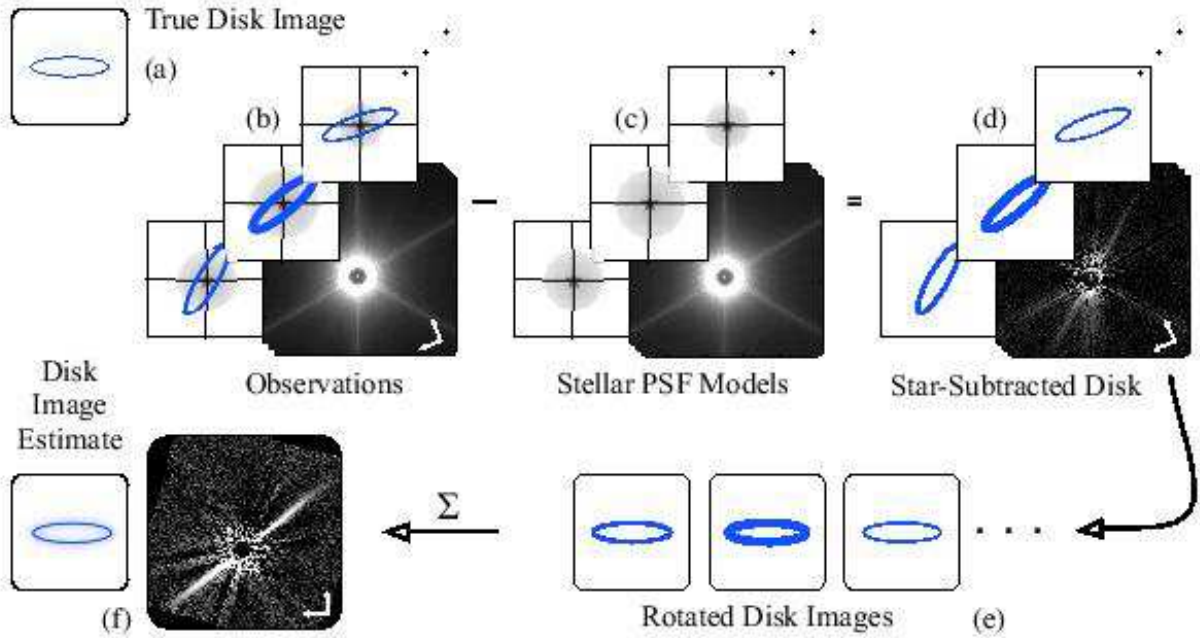}
\caption{The roll subtraction process, which estimates the image of the disk in the presence of a fluctuating stellar PSF.  The disk \textit{(a)} is observed on an alt-az telescope with no image de-rotation \textit{(b)}.  The PSF is fixed relative to the detector while the disk image rotates due to sidereal motion.  The observed disk images are blurred by different amounts by the time-varying PSF, illustrated by different line weights the disk images and the size of the circular stellar halo.  Estimates of the PSF scaled and positioned to match the stellar image \textit{(c)} are subtracted from the observations, leaving residual disk images \textit{(d)}.  These images are transformed to a common frame \textit{(e)} and combined to estimate the disk image \textit{(f)}.}\label{fig:subtraction_overview}
\end{figure}

We developed an algorithm suited to the self-subtraction of time-varying AO PSFs and applied it to the reduction of our near-IR images.  In essence, because the PSF remains in a fixed orientation while the edge-on disk rotates, one may use PSF information spanning several exposures to estimate the stellar PSF (effectively removing the disk).  Our refinement to the procedure is to estimate the PSF for each exposure, rather than for the ensemble.  The PSF estimates are then used to subtract the stellar image from each frame.  The residuals are transformed to a common sky orientation and then combined to estimate the star-subtracted object field (Fig.~\ref{fig:subtraction_overview}), corresponding to the disk emission.  We detail our procedure, including the refinements for tracking changes in the time-variable AO PSF, in Appendix~\ref{ap:roll_sub}.

\subsection{Calibration}\label{subsec:calibration}

Immediately after bias subtraction and flat fielding, we corrected the camera's geometric distortion in each image.
However, no attempt was made to calibrate for the scale and orientation of the detector on the sky, as any differences from nominal values are expected to be minor.
When visually comparing with \textit{HST} imaging~(\S\ref{subsec:hst_data}), we found a $0\fdg 1$ rotation between the two sets.

Our calibrations for photometry require measurement of zero point and PSF.  We bracketed our observations of \AUMic\ with photometric standard stars SJ\,9170 and HD\,205772 on the first night, and GJ\,811.1 on the second~\citep{elias_etal82, persson_etal98}.  These observations were used to determine the photometric zero point in each band.  The stars were positioned outside the coronagraphic spot, and aperture photometry was used to measure the stellar brightness.  Since large aperture radii could be used on the well-exposed stellar images, no encircled energy corrections were used in determining zero points.  For all measurements, we applied an airmass correction assuming extinction values appropriate for Mauna Kea~\citep{krisciunas_etal87}.

To measure the scattering properties of the dust, we express the disk flux relative to that of the star.  However, measuring the brightness of \AUMic\ is a challenge with this instrumental configuration because it is too bright for direct unocculted imaging, while observations using the focal plane mask complicate calibration.
Here we adopt 2MASS photometry for the stellar brightness~\citep[$J = 5.436\pm 0.017$, $H = 4.831\pm 0.016$, $K = 4.529 \pm 0.020$;][]{skrutskie_etal06}.
We ignore the small ($\sim$ few centi-mag) color correction when transforming from \bK\ to \bKp, as we expect disk photometry to be dominated by other errors.
To determine the brightness of the star in \bLp, we first measured the brightness of GJ\,811.1 through the partially-transmissive focal plane spot in \bL.  We then applied a color correction to compute the zero point (behind the spot) in \bLp, and measured the brightness of the \AUMic\ using the same photometric aperture, finding 4.38\,$\pm$\,0.04 mag in \bLp.
We note that our attempts to express the relative disk and stellar flux ratio should take \AUMic's variability into consideration.
The star regularly flares in the X-ray and EUV regions of the spectrum~\citep[for a review with application to grain dynamics, see][]{augereau&beust06}.
Periodic variations in the visible regions of the spectrum, thought to be caused by spots~\citep[$\Delta m_V\sim 0.35$\,mag, $P=4.865$\,d;][]{torres&ferrazmello73, cutispoto_etal03}, are also relevant.
However, as noted by~\citetalias{metchev_etal05}, for \AUMic\ these are likely not problematic since the contrast between spots and the photosphere is lower in the near IR.

It is convenient to measure surface brightness in rectangular photometric apertures, as are used in the calculations detailed in~\S\ref{subsec:sb_prof}.  Because of the difficulty in using coronagraphic observations to infer the off-axis (unobscured) PSF, we apply an aperture correction based on the enclosed energy in this aperture by using the unobscured observations of photometric standard stars as PSF references.  In each band, we average the enclosed energy over azimuthal rotation of the reference PSF to simulate the final roll-subtracted image's PSF.
The aperture corrections we derived from these enclosed energy measurements are susceptible to variability in the AO PSF because the reference PSFs are measured non-contemporaneously from those of \AUMic.  To estimate the systematic errors in the overall flux levels, we examined the random error in the enclosed energy of the reference PSF exposures in each band.  The largest fluctuation on a single night was 16\%.  Although the combination of data from both nights will reduce this uncertainty ($\lesssim 11\%$), the potential for unmeasured changes in the PSF from the reference star to science measurements remains.  With caution in mind, we estimate the uncertainty in flux calibration error in the final near-IR images at 20\% and note that, in general, unsensed fluctuations will tend to affect the shorter-wavelength observations to a greater degree.
This systematic calibration uncertainty will affect the absolute levels of the surface brightness profiles we measure in~\S\ref{subsec:sb_prof}, and depending on the degree in correlation between errors in calibration in different bands, the disk colors~(\S\ref{subsec:scat_col}).  Measurements of color gradients will not be affected by this type of error.

\subsection{HST Imaging}\label{subsec:hst_data}

In order to compare disk images over a wider range of wavelengths, we reduced and analyzed data previously obtained with the \textit{Hubble Space Telescope} Advanced Camera for Surveys in the \bFW\ filter ($\lambda_c$=606\,nm, $\Delta\lambda$ = 234\,nm) on 2004 April 03~\citepalias{krist_etal05}.  We acquired the flatfield images of \AUMic\ and the PSF reference star HD\,216149 from the \textit{HST} \textsc{OPUS} pipeline.  These data were further calibrated by dividing by an appropriate spotflat and multiplying by a pixel area map.  A final image of \AUMic\ was constructed by averaging three frames of 750 seconds integration each, with appropriate filtering to reject cosmic rays.  The final image of HD\,216149 was constructed by averaging eight frames of 225 seconds each.  We subtracted the PSF subtraction in a manner described by~\citetalias{krist_etal05}.  There was no need to apply our specialized roll-subtraction technique (Appendix~\ref{ap:roll_sub}) to these data because of the stability of the ACS PSF.  The PSF-subtracted \AUMic\ image was then corrected for geometric distortion using a custom \textsc{IDL} routine (J. Krist, private communication).  We used the Tiny Tim PSF model\footnote{\url{http://www.stsci.edu/software/tinytim/}}~\citep{krist&hook04} to compute the enclosed energy in rectangular apertures to enable comparison of surface brightness profiles with our near-IR data.

We used the same method for computing midplane surface brightness profiles described previously~(\S\ref{subsec:calibration}).  We checked the consistency of our \bFW\ midplane surface brightness measurements against those reported in~\citetalias{krist_etal05}'s Fig.~4 (using their 0\farcs 25\,$\times$\,0\farcs 25 apertures) and found that our measurements of the same data were uniformly 0.20\,$\pm$\,0.05\,mag brighter.  The discrepancy is due to the fact that~\citetalias{krist_etal05} divided their images by a factor of 1.124 when correcting their pixel areas from 28\,$\times$\,25\,mas\,pixel$^{-1}$ to 25\,$\times$\,25\,mas\,pixel$^{-1}$ (J. Krist, private communication).  However, the photometric calibrations, both in the image headers and produced by \textsc{SYNPHOT}, assume that the final undistorted image is processed using \textsc{DRIZZLE}.  The drizzled data product is similar to the manual calibration performed above and in~\citetalias{krist_etal05}, except that it does not include a uniform scaling of the image by a factor of 1.124.  To correct the error, the disk brightness values reported in~\citetalias{krist_etal05} should be increased by this factor.  We adopted~\citetalias{krist_etal05}'s value of $8.63\pm 0.03$\,mag for the stellar brightness.  Aperture size aside, our measurement methodology of the \bFW\ midplane surface brightness profile presented in~\S\ref{subsec:sb_prof} produces results consistent with~\citetalias{krist_etal05} after scaling their surface brightness by a factor of 1.124.
The stability of the \textit{HST}/ACS PSF ensures that the flux calibration uncertainty, set by the 0.03\,mag uncertainty in stellar brightness, is much lower than those of the near-IR bands.  The stellar brightness in \bFW\ was measured in unocculted images at the same epoch, so our measurements are unaffected by the star's variability.

\section{RESULTS}\label{sec:results}

\subsection{Disk Morphology}\label{subsec:structure}

\begin{figure}
\plotone{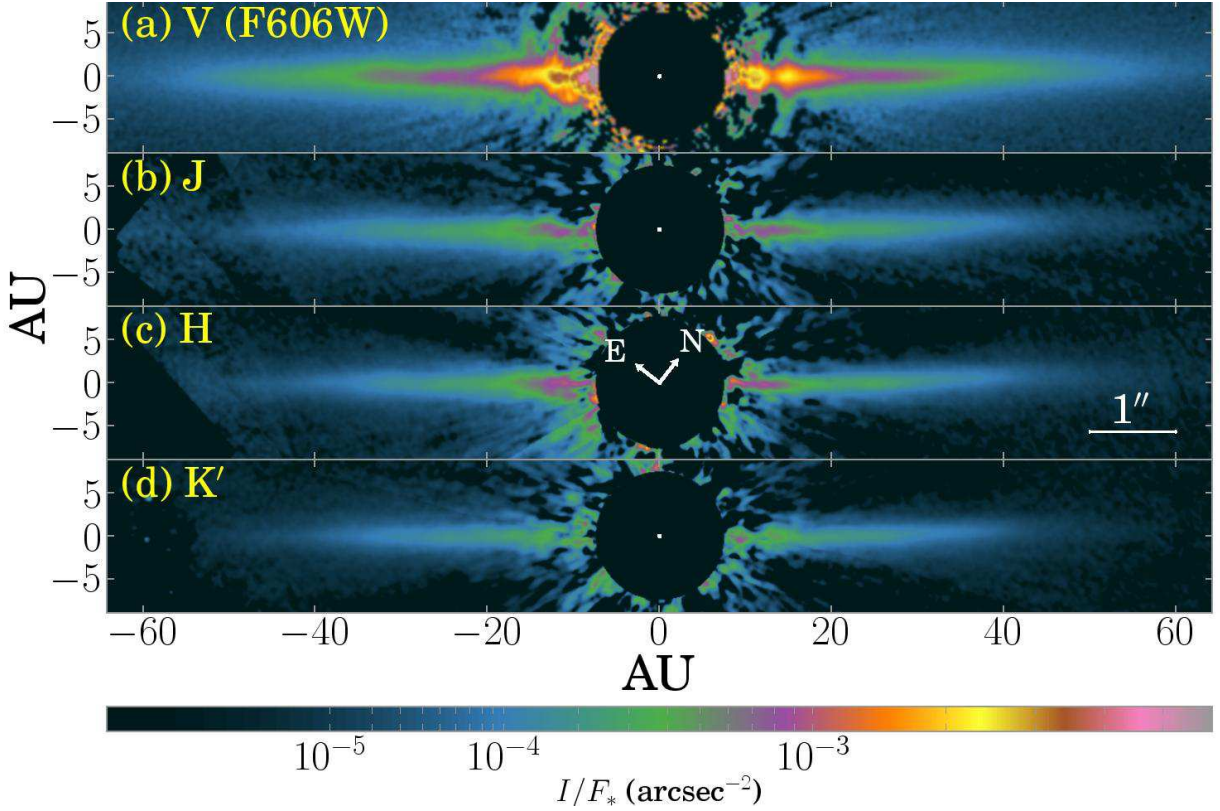}
\caption{Images of the \AUMic\ debris disk with direct starlight removed via PSF roll subtraction~(\S\ref{subsec:psf_sub}).  Data are displayed on a square-root scale, and in each band the disk brightness $I$ is divided by the stellar flux $F_*$, allowing for comparison of color differences intrinsic to the dust rather than the star.  In each image, a circular software mask (1\farcs 5 diameter) is applied to obscure subtraction residuals about the stellar location, marked by a small white circle.  The near-IR images have been additionally smoothed by a Gaussian matched to the resolution of the PSF to eliminate small high-frequency errors introduced by the masking process (Appendix~\ref{ap:roll_sub}).  The data in panel \textit{(a)} were obtained with the ACS coronagraph aboard \textit{HST}, while the data in \textit{(b)--(d)} are newly acquired via Keck AO.  The blue color of the dust is indicated by the trend of decreasing scattering efficiency toward longer wavelengths in panels \textit{(a)--(d)}.}\label{fig:images}
\end{figure}

\begin{figure}
\plotone{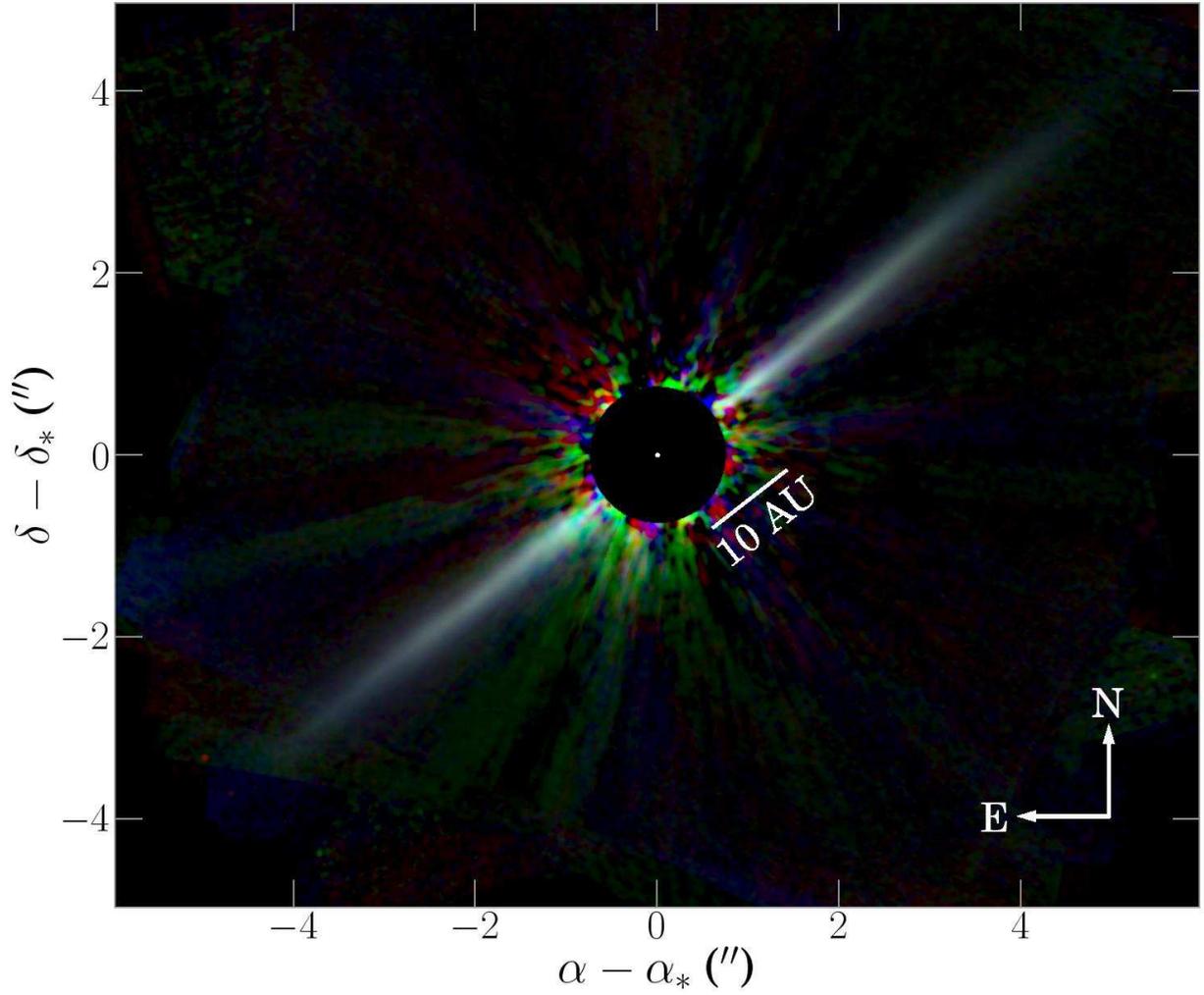}
\caption{A \bJ\bH\bKp\ composite image using the data in Figure~\ref{fig:images}.  The data are displayed relative to the stellar brightness, highlighting the intrinsic color of the dust.  The structures emanating from the mask outside of the disk are residuals from the stellar PSF subtraction.  The blue color of the disk is visible, as is evidence for substructure.}\label{fig:imcomposite}
\end{figure}

The PSF-subtracted images are shown in Figures~\ref{fig:images} and~\ref{fig:imcomposite}.  As observed by other authors, the disk has a near-edge-on morphology.  The data in each band of Figs.~\ref{fig:images} and~\ref{fig:imcomposite} has been divided by the corresponding stellar brightness, such that the resulting colors trace the relative scattering efficiencies of the dust.  We find the disk decreases in brightness relative to the star with increasing wavelength, up to \bKp-band, indicating an overall blue color.  We do not detect the disk with strong confidence in \bLp.

\begin{figure}
\plotone{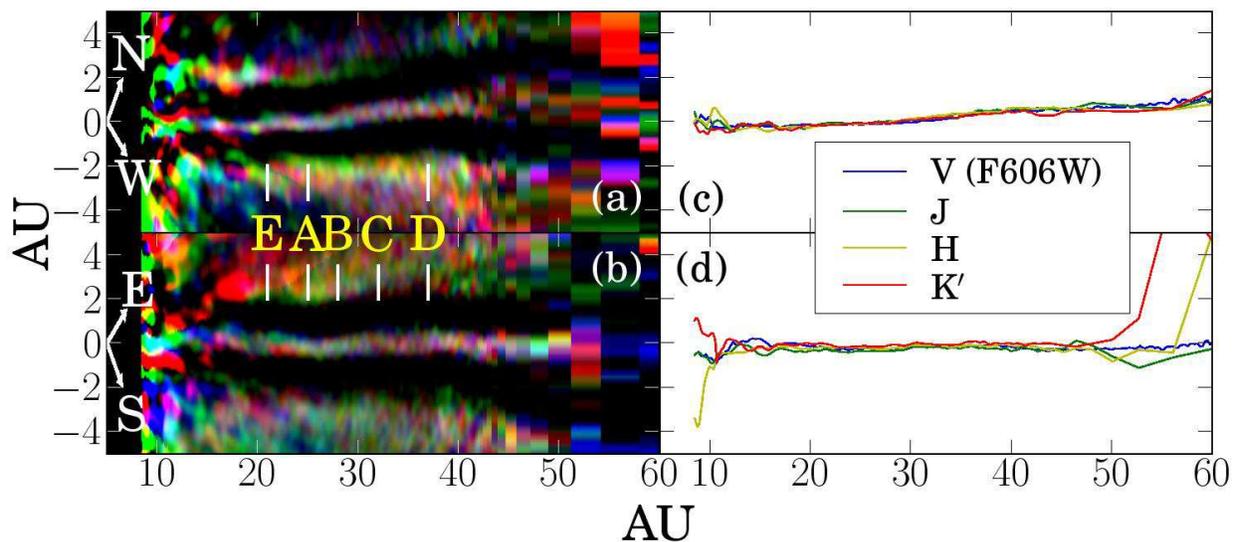}
\caption{Vertical structure in the near-IR disk images.
Panels \textit{(a)} and \textit{(b)} show a \bJ\bH\bKp\ composite of disk images, whereby a Gaussian fit to the vertical structure of the disk has been subtracted.  The residuals clearly show the location of the sharp midplane, and have been scaled by a smooth fit to the amplitudes of the Gaussian functions.  The SE side is flipped about the star to allow for direct comparison to the NW.  The locations of physical features are indicated (A--E; cf. Table~\ref{tab:struct}).
Panels \textit{(c)} and \textit{(d)} plot the variation of the vertical midplane position resulting from Gaussian fits to the vertical profile.  No significant differences in midplane position with wavelength are seen.}\label{fig:vertstructure}
\end{figure}

The disk midplane shows vertical structure (Fig.~\ref{fig:vertstructure}).
We fit a Gaussian function to the vertical profile of the disk as a function of projected distance.  We subtract this profile from the image, scale the residuals by a smooth fit to the amplitudes of the Gaussians, and display the results in the left panels of the Figure.  This procedure clearly reveals the location of a sharp midplane.
Broader features also show variation --- the width of the best-fit Gaussians is not constant with projected distance from the star.  This effect has been noted and studied by other authors~\citep[\citetalias{krist_etal05};][]{graham_etal07}.  Capturing the vertical structure of the disk is important for detailed physical models of the system.  However, here and in~\S\ref{sec:analysis}, we restrict ourselves to considering only the most basic disk properties and defer detailed two-dimensional modeling for future work.  We will discuss the disk substructure (brightness enhancements and deficits) in~\S\ref{subsec:substructure}.

\subsection{Surface Brightness Profiles}\label{subsec:sb_prof}

\begin{figure}
\centering\includegraphics[width=.7\textwidth]{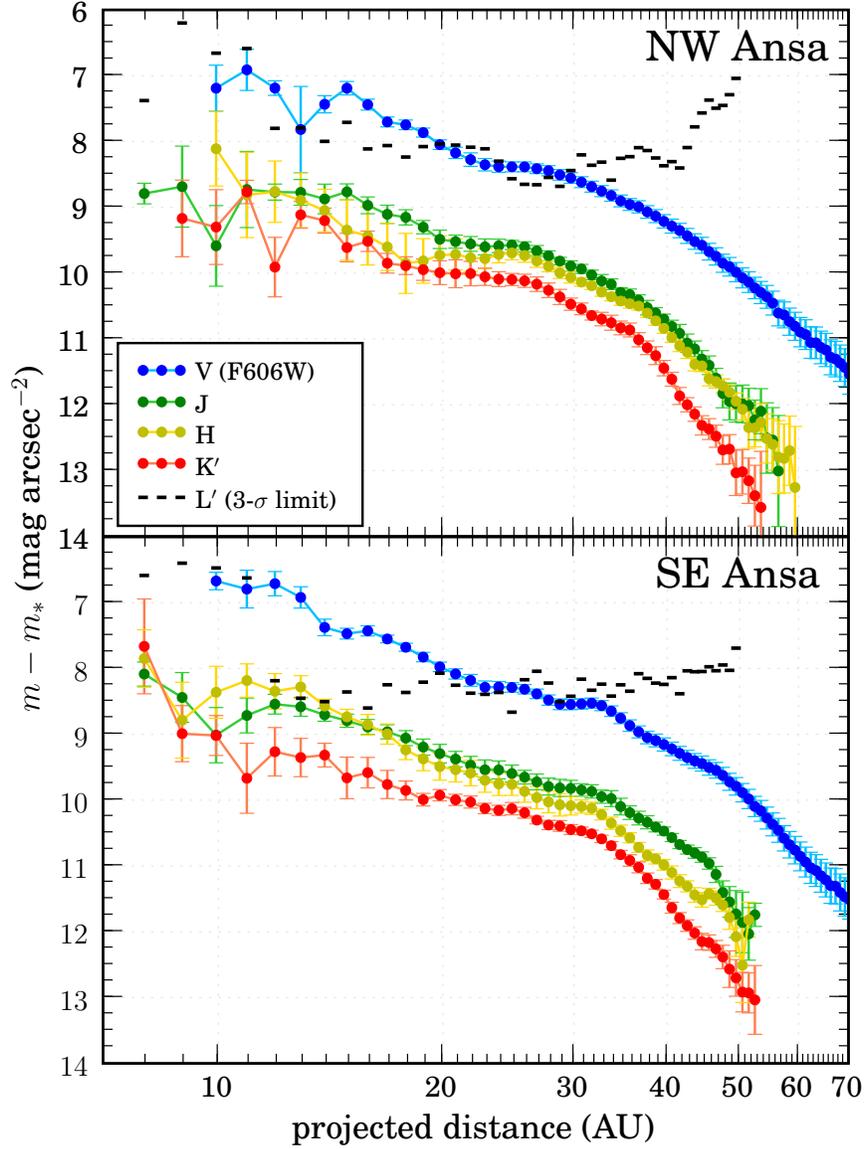}
\caption{Midplane surface brightness profiles for the disk ansae.  Flux was gathered in 0\farcs 1$\,\times\,$0\farcs 5 apertures, which were placed according to the spline fit to the vertical disk midplane.  The measurements are shown relative to the stellar brightness to highlight the intrinsic scattering properties of the dust in the \bJ- to \bKp-bands.  The uncertainties represent 1-$\sigma$ random errors, and do not include systematic errors in calibration~(\S\ref{subsec:calibration}).  We also indicate 3-$\sigma$ upper limits to the \bLp\ brightness.}\label{fig:sbprof}
\end{figure}

Midplane surface brightness profiles are useful metrics of disk structure.  These profiles average over the vertical extent of the observed scattered light image, which is naturally integrated along the line of sight.  This facilitates comparison with scattered light models by reducing the brightness distribution to one dimension.
We note that this convenient technique can fail to capture variation in the projected vertical extent of the disk, though
we do follow the vertical centroid as a function of projected position.  Prior to computing our profiles, we fit a spline to the vertical midplane position along each ansa.  We used 0\farcs 1\,$\times$\,0\farcs 5 photometric apertures centered on these positions in unsmoothed images to extract the photometry (Figure~\ref{fig:sbprof}).  The width of this aperture is chosen to provide sufficient spatial resolution along the disk midplane, while the height is sufficient to capture several vertical FWHM of the inner disk.  As reported by~\citet{graham_etal07}, the apparent \bFW\ disk thickness increases outward, reaching the 0\farcs5 aperture height at $\sim$\,60\,AU.  Beyond this point, these photometric apertures do not perform a vertical average; rather, they sample the midplane brightness.

Our PSF-subtracted \bLp\ data exhibit very low S/N structures along the disk plane in \bLp\ at separations of $\sim$\,1\arcsec\ from the star, though these may result from PSF subtraction errors.  In Figure~\ref{fig:sbprof}, we show upper limits on the disk brightness in this band.

The uncertainties in Fig.~\ref{fig:sbprof} represent the random measurement errors.  At each radius, the contribution from residual speckle and background noise sources was estimated from the standard deviation of photometry in apertures placed in an annulus, excluding the locations near the disk.  The photon noise from disk photometry is also factored into the random errors.
The inner and outer edges of the profiles are set by the requirement that $\mathrm{S/N}\ge 1$.
To estimate the contribution of PSF subtraction to the systematic errors, at each radius we compared the mean values of the off-disk apertures (placed in an annulus) to the random error estimated from their standard deviation.  These are generally less than one, suggesting that any bias in the profiles is smaller than the estimated random errors.
Another possible mechanism for systematic error arises in the near-IR data because the PSF estimates are derived from the target exposures~(\S\ref{subsec:psf_sub} and Appx.~\ref{ap:roll_sub}).  If there is insufficient field rotation in the images, light from the inner disk may be subsumed in the PSF estimate, resulting in a self-subtraction of the disk.  Nearly all of our datasets have $\Delta\mathrm{PA}>30\degr$ (Table~\ref{tab:obs}), sufficient to exclude this as a source of significant error.  However, the 2004 Aug 29 \bJ-band data have $\Delta\mathrm{PA}\simeq 17\degr$, which corresponds to one disk midplane FWHM at $\sim 8.5$\,AU.  The 2004 Aug 30 data have much more field rotation, so the effect on the inner region of the composite \bJ-band surface brightness profiles is likely to be small.
Finally, the uncertainties in Fig.~\ref{fig:sbprof} do not include potential errors due to flux calibration, which may be as large as 0.2\,mag for the \bJ\bH\bKp\ bands~(\S\ref{subsec:calibration}).
We conclude that, relative to the random errors, the systematic errors are unlikely to affect the shape of the surface brightness profiles at separations greater than 8.5\,AU.

By comparing our near-IR data to those of~\citetalias{liu04} and~\citetalias{metchev_etal05}, we confirm a break in the overall midplane surface brightness profile around 30--35\,AU.
A break at $\sim$\,15\,AU in the \bFW\ data was reported by \citetalias{krist_etal05}, though the innermost points in our near-IR profiles are consistent with both a slight flattening and no break at all.

Broken power laws provide a compact description of the observed surface brightness profiles.  Using flux-based units rather than magnitudes, we fit a function $f(b)\propto b^{-\alpha}$ to the midplane surface brightness profiles in Fig.~\ref{fig:sbprof}, and report the resulting indices $\alpha$ in Table~\ref{tab:power_laws}.  We scale the formal 1-$\sigma$ random errors by $\sqrt{\chi^2_\nu}$ when this quantity is $>1$ to partially account for the ill-fittedness of a strict power-law to profiles which exhibit substructure.  Errors in subtraction may be correlated because some PSF structures have significant radial extent.  We ignore measurement covariance in the analysis, and thus potentially underestimate $\chi^2_\nu$ and the quoted errors.
We chose to fit over two domains in projected separation: an inner region of 15.0--32\,AU, and an outer region of 32--60.0\,AU.  The innermost boundary was chosen to mitigate possible biases from systematic errors.  The outermost boundary is an upper limit, and is more precisely set by the availability of data in Fig.~\ref{fig:sbprof}.  The break between the two regions was chosen to correspond to the kink in the midplane surface brightness profile of the SE ansa, corresponding to feature C~(\S\ref{subsec:substructure}).
The power-law fits of our \bH-band data are somewhat consistent with those of~\citetalias{liu04}, who measures for the NW (SE) ansa $\alpha=1.4\pm 0.3$ $(1.0\pm 0.3)$ over 20--35\,AU and $4.4\pm 0.3$ $(4.4\pm 0.4)$ over 35--60\,AU.  Using the same method as used in Table~\ref{tab:power_laws} on these ranges (but different aperture sizes and positions), we obtain $1.6\pm 0.2$ $(1.5\pm 0.1)$ and $4.4\pm 0.2$ $(3.4\pm 0.2)$, respectively.  The values we compute for the SE slopes are steeper in the inner region, but shallower outside.
Our measurements are consistent with those of~\citetalias{metchev_etal05} over similar domains.  Those authors measure $\alpha=1.2\pm 0.3$ over 17--33\,AU and $4.0\pm 0.6$ over 33--60\,AU.  Using these ranges and averaging over ansae (but different aperture parameters) we obtain $1.4\pm 0.2$ and $3.8\pm 0.2$.
We have significant differences in our \bFW\ power-law indices relative to those of~\citetalias{krist_etal05}.  The authors, using apertures that are 0\farcs 25 tall in the direction perpendicular to the midplane, calculate $\alpha=1.8$ over the domain 15--43\,AU, and $4.7$ in the region beyond 43\,AU, both of which are somewhat steeper than our findings of $1.6\pm 0.1$ and $4.1\pm 0.2$ (measuring from 15--43\,AU and 43--70\,AU with our apertures).  However, measuring with 0\farcs 1\,$\times$\,0\farcs 25 apertures over the region 43--60\,AU and averaging the fit indices over both ansae, we find $\alpha=4.7$ --- consistent with~\citetalias{krist_etal05}.
The differences in these comparisons serve to underscore the sensitivity of such power-law measurements to methodology.  These differences may arise from PSF subtraction residuals, aperture sizes, aperture center locations, and the fit domain.  Inferences between different datasets should be based on consistent power-law fits.

\subsection{Disk Color Variation}\label{subsec:scat_col}

\begin{figure}
\plotone{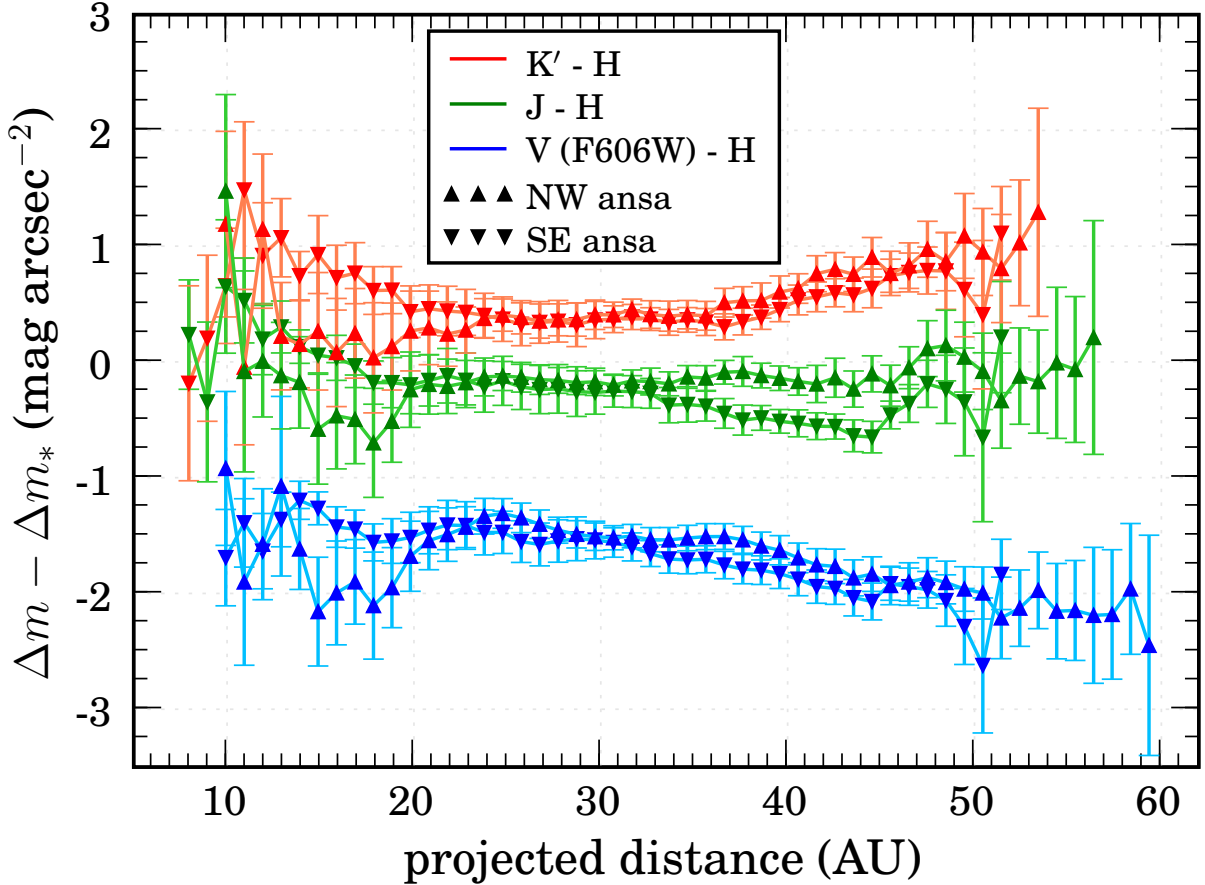}
\caption{The disk color vs. the projected distance along the disk midplane.  We compute disk colors relative to the \bH-band midplane surface brightness profile.  The contribution of the intrinsic stellar color has been removed.  The disk appears blue between the visible and near-IR and between \bH\ and \bKp.  Note that \bKp-\bH\ colors are shown, which accounts for the reversed gradient from \bFW-\bH.  Systematic uncertainties in calibration are not included, though these are expected to be $\lesssim 0.3$\,mag for \bKp-\bH\ and \bJ-\bH, and $\lesssim 0.2$\,mag for \bFW-\bH~(\S\ref{subsec:calibration}--\ref{subsec:hst_data}).  In the inner disk ($\lesssim 35$\,AU), the colors are consistent with a flat profile.  The outward branching of these curves in the outer disk indicates blue color gradients (most clearly indicated by the \bFW-\bKp\ color).  These gradients may indicate differences in grain sizes or compositions.}\label{fig:colorgrad}
\end{figure}

The blue color, relative to the star, of the scattered light was first reported by~\citetalias{krist_etal05}.  It becomes increasingly blue at larger radii (\textit{F435W}-\textit{F814W}\,=\,0.2--0.5\,mag from 30--60\,AU).  The disk also scatters blue when comparing visible data to the \bH-band measurements of~\citetalias{liu04} and~\citetalias{metchev_etal05}.  The blue color of the scattered light extends to our \bJ\bH\bKp\ observations.  Figure~\ref{fig:colorgrad} shows the color as a function of projected radial position along the disk midplane.  The stellar contribution to the apparent disk color has been removed in order to highlight intrinsic grain scattering processes.  Systematic uncertainties in calibration are not included, though as discussed in~\S\ref{subsec:calibration}--\ref{subsec:hst_data}, these are expected to be $\lesssim 0.3$\,mag for \bKp-\bH\ and \bJ-\bH, and $\lesssim 0.2$\,mag for \bFW-\bH.  In the inner disk ($\lesssim 35$\,AU), the observed color gradient is consistent with a flat profile, while the outer disk is increasingly blue with stellocentric distance.

\subsection{Disk Substructure}\label{subsec:substructure}

\begin{figure}
\plotone{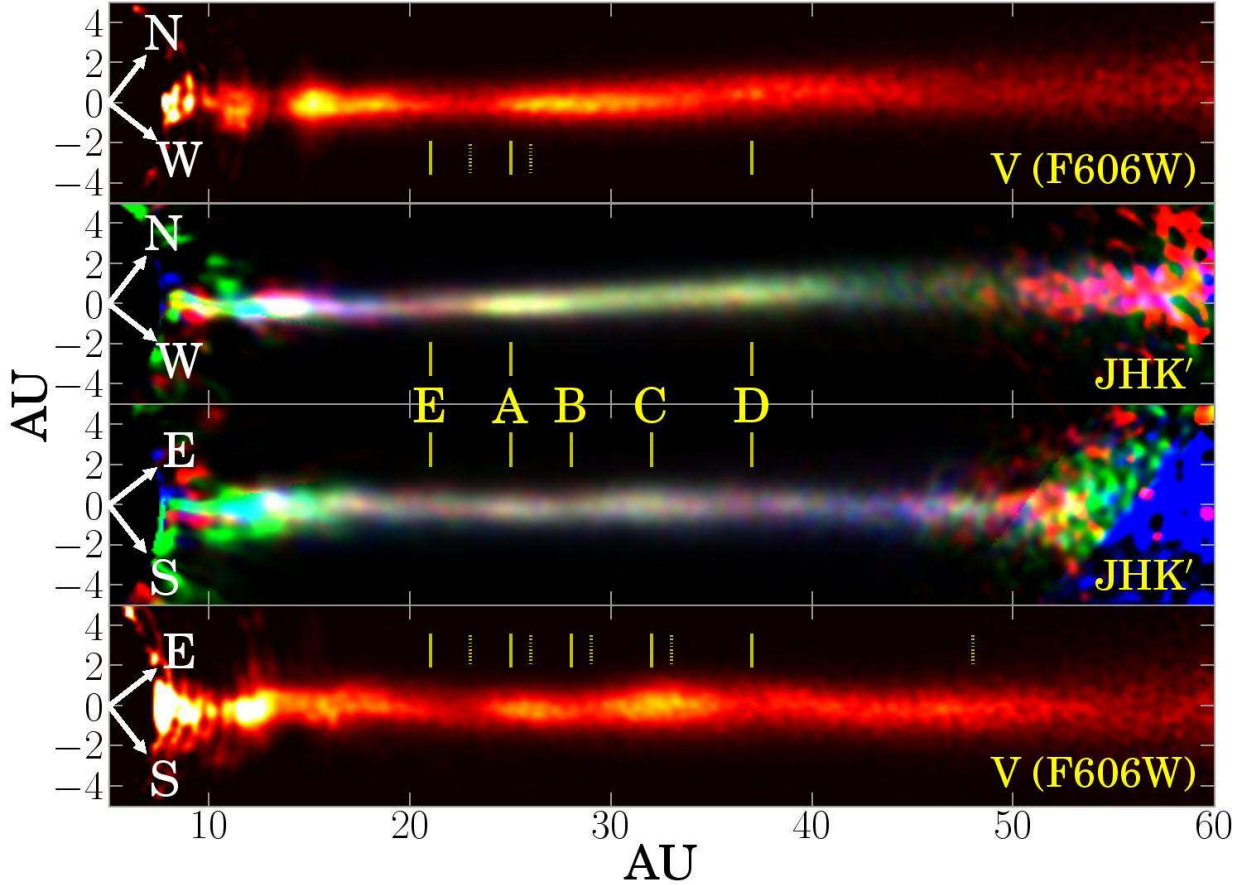}
\caption{Substructure in the disk images.  The upper and lower panels are \bFW\ data, while the middle panels are \bJ\bH\bKp\ composites.  The image in each band is scaled by a spline fit to an average of its NW and SE midplane surface brightness profiles~(\S\ref{subsec:sb_prof}).  The SE side is flipped about the star to allow for direct comparison to the NW.  Feature locations in the near-IR are indicated by solid lines (A--E), while the original identification of feature locations in \bFW\ data by~\citetalias{krist_etal05} are shown as dotted lines (cf. Table~\ref{tab:struct}).  A broad clump in the SE ansa at 48\,AU was also identified by~\citetalias{krist_etal05}.  We confirm the presence of a brightness deficit at location E.  In our favored models (\S\ref{sec:analysis}), the bulk of the scattered light comes from small grains outside of 40\,AU.  Therefore, the features at projected distances inward of 40\,AU (A--E) must arise from azimuthal perturbations in the dust distribution outside this radius.  The origin of the substructures is unknown; they may result from the gravitational influence of unseen planets.}\label{fig:substructure}
\end{figure}

\begin{figure}
\plotone{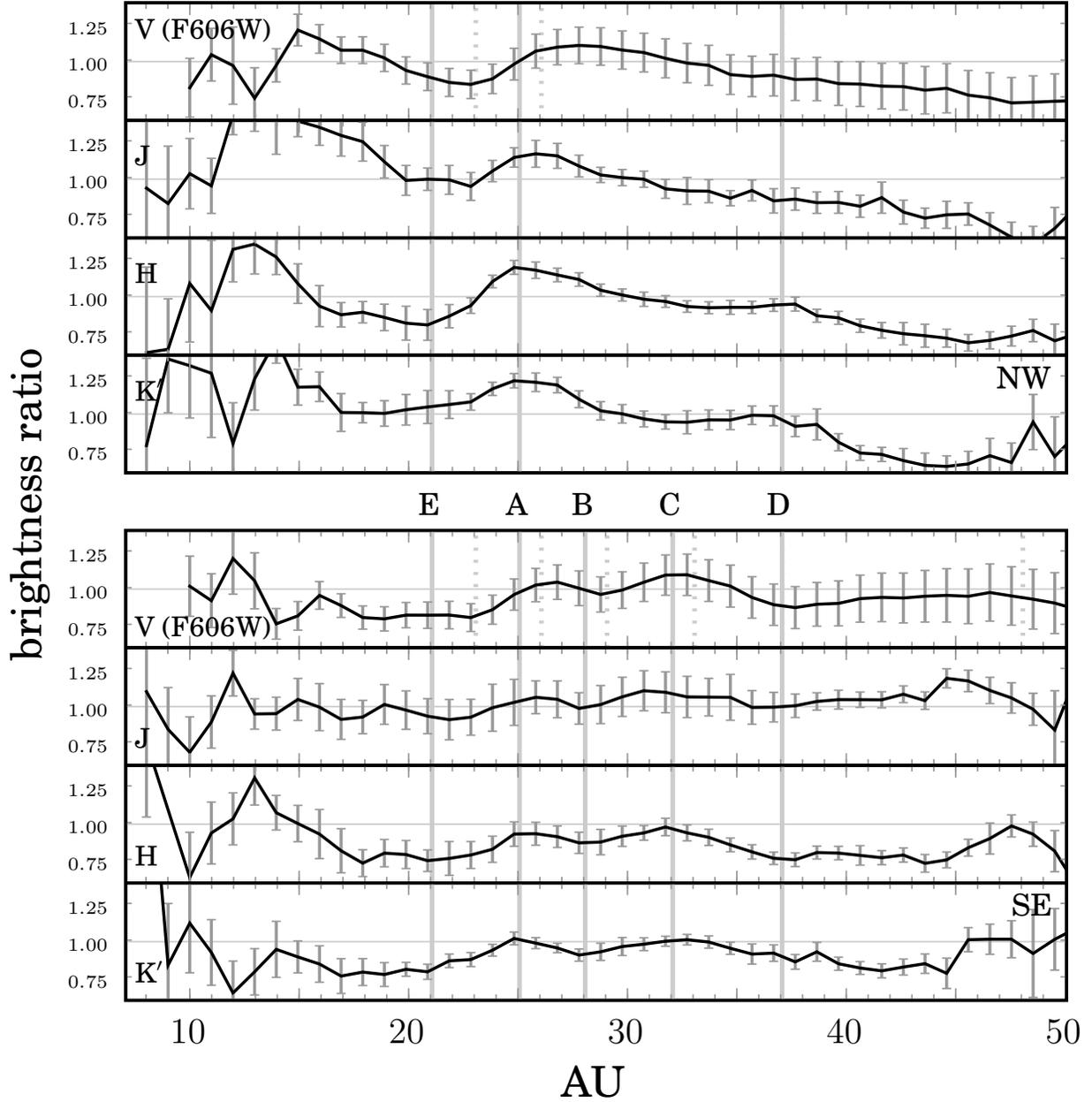}
\caption{Substructure in disk profiles.  Surface brightness profiles computed with 0\farcs1\,$\times$\,0\farcs1 apertures have been processed in the same manner as Fig.~\ref{fig:substructure}, by dividing each wavelength's profile by a smooth spline function derived from an average of the NW and SE profiles from Fig.~\ref{fig:sbprof}.  Feature locations derived for the near-IR data are indicated by solid vertical lines (A--E), while dotted lines indicate the feature locations in \bFW\ data by~\citetalias{krist_etal05}.}\label{fig:substructure_profs}
\end{figure}

The \AUMic\ debris disk is known to exhibit non-uniformity in its midplane surface brightness distribution~\citepalias{liu04, krist_etal05, metchev_etal05}.
In order to remove global structure and highlight localized variations, we divided each brightness distribution by a smooth function in each band.  We obtained weighted fits of spline functions to the NW and SE ansae's midplane surface brightness profiles~(\S\ref{subsec:sb_prof}) and averaged the results.  The order and smoothness of the splines used to enhance the data have direct bearing on the spatial frequency content of the resulting map and may affect the observed positions and brightness of disk features.  We used the \texttt{curfit} routine of \textsc{FITPACK}\footnote{See \url{http://www.netlib.org/dierckx/}.} to fit a cubic spline to the data in Fig.~\ref{fig:sbprof}, with smoothness $s=m+\sqrt{2m}$, where $m$ is the number of data points in the profile.
Figure~\ref{fig:substructure} shows an \bFW\ image and a \bJ\bH\bKp\ color composite that have been processed in this manner.  The annotations give the locations of substructural features (A--E).  To allow for more quantitative comparison of disk features, in Figure~\ref{fig:substructure_profs} we show surface brightness profiles processed in the same manner.  In order to highlight the structure of midplane brightness, these profiles are computed with 0\farcs1\,$\times$\,0\farcs1 apertures.  

The imperfect removal of time-variable features in the PSF may introduce systematic errors which can masquerade as disk structures, underscoring the value of independent observations when identifying particular features.  \citet{liu04} identified several features of the disk substructure, which were also observed by~\citetalias{krist_etal05} and~\citetalias{metchev_etal05}.
In independent observations such as these, any changes seen in the structures may be due to PSF-subtraction artifacts.  However, given that \AUMic\ exhibits brightness variability due to starspots, it is also possible that these spots differentially illuminate the disk such that the spatial distribution of disk brightness is also time-variabile.  We also note that differences in feature position and brightness from different authors may arise via analysis methodology.
As we have done in Figs.~\ref{fig:substructure} \& \ref{fig:substructure_profs}, it is desirable to highlight features by dividing the data by a function $f(b)$, where $b$ is the projected stellar separation.  For example,~\citetalias{liu04} used $f(b)=\vert b\vert^{-1}$ (see his Fig.~3) and~\citetalias{metchev_etal05} used $f(b)=b^{-2}$ (their Fig.~3b).  In contrast,~\citetalias{krist_etal05} use a fourth-order polynomial fit to the surface brightness profile for $f(b)$, while we use a spline fit (detailed above).  These more complicated functions can better capture the overall brightness profile of the disk, which does not follow a single power law.  On the other hand, they can suppress the appearance of the broadest features (e.g. feature D in the NW ansa).

The application of the spline fit to the \bFW- and \bJ\bH\bKp-band data allows us to eliminate analysis methodology as a potential source of variation in feature brightness and position.  While there are minor differences in the apparent characteristics of features between the near-IR bands, distinct differences in feature location arise when comparing those seen in visible to the near-IR.

Feature A, which is a brightness enhancement seen in both ansae, is $\sim 1$\,AU further from the star in the visible (26\,AU) compared to the near-IR (25\,AU).  There is also evidence for a change in the position of this feature in the near-IR; in particular, the \bJ-band centroid is further out than in \bH\ or \bKp\ (cf. Fig.~\ref{fig:substructure_profs}).
Feature B is a brightness deficit in the SE ansa that also exhibits a similar inward-moving centroid with increasing wavelength.
Feature C is a SE brightness enhancement, broader than A, that shows a different position depending on the waveband.  In \bJ\ and \bH, the feature centroid is at $\sim32$\,AU, while it peaks closer to 33\,AU in \bFW\ and \bKp.
Feature D (37\,AU) corresponds to a broad enhancement on the NW ansa, a narrower dip in the SE ansa brightness, and the location of the vertical displacement of the NW midplane.  The position of this feature is not significantly wavelength-dependent in these data.
We confirm~\citetalias{krist_etal05}'s identification of a dip in midplane brightness inward of feature A, which we denote `E.'
The locations of the other features seen in \bFW\ agree with the positions given by~\citetalias{krist_etal05}.  The \bH-band positions and characteristics of features A--D compare favorably to the results of~\citetalias{liu04}.  Together these strengthen the evidence for $\sim 1$\,AU differences in the positions of features A--C.  In contrast,~\citetalias{metchev_etal05} measures positions for A and B slightly closer to the star (by 2--3\,AU); however those data are of lower S/N and the differences may not be significant.
A summary of these results is given in Table~\ref{tab:struct}.

There are additional deviations in brightness that are detected in multiple bands.
The broad enhancement at $\sim 46$\,AU in the SE noted by~\citetalias{krist_etal05} is also detected in all the bands we consider.
The feature seen in the \bFW\ data, characterized by a peak enhancement at 15\,AU and sharper drop at 12\,AU, has corresponding characteristics in the near-IR bands.  In \bJ\bH\bKp, the peak is seen at 12--13\,AU and the inner cutoff at 11--12\,AU.  The outer extent of this feature decreases gradually from $\sim 20$\,AU in \bFW\ to 17\,AU in \bKp.
The peak enhancement in the SE at $\sim $12\,AU in \bFW\ also has corresponding enhancements at 12--13\,AU in \bJ\ and \bH; an examination of the \bKp-band surface brightness profile (Fig.~\ref{fig:sbprof}) suggests the enhancement may also be present at $\sim 13$\,AU in this band, but is not visible in Figs.~\ref{fig:substructure} \& \ref{fig:substructure_profs} due to the spline fitting process.

\subsection{Point Source Detection Sensitivity}\label{subsec:pt_src}

We do not detect any point-like sources in the disk midplane.  We have developed a methodology for utilizing the artificial insertion of point sources into an image to measure the detection sensitivities in both the background and the disk midplane.  We detail our technique in Appendix~\ref{ap:pt_src}, and present our sensitivity limits in Figure~\ref{fig:sensitivity}.
We indicate the predicted brightnesses of model giant planets from~\citet{burrows_etal97}, placing limits on the presence of young, massive companions.  We note the caveat that at young ages, the emission from Jupiter-mass planets may be sensitive to initial conditions~\citep[e.g.][]{fortney_etal05, marley_etal07}.

\section{ANALYSIS}\label{sec:analysis}

Models of the grain size and space distributions that reproduce the scattered light and thermal emission probe the dynamical processes affecting the dust distribution.
Previous models of the \AUMic\ disk have attempted to model the SED~\citep{chen_etal05} and the scattered light profiles~\citepalias[e.g.][]{krist_etal05}. 
\citet{metchev_etal05},~\citet{augereau&beust06}, and~\citetalias{strubbe&chiang06} tackled the task of simultaneously fitting both.
Using continuous power-law descriptions for the grains' space and size distributions (sub-\micron\ to mm sizes),~\citetalias{metchev_etal05} find they cannot simultaneously account for the scattered-light color, the SED, and the break in the surface brightness profiles at $\sim$\,35\,AU.
As an alternative to power-law density distributions,~\citet{augereau&beust06} obtained surface density models via scattered-light profile inversion.  They tuned the size distribution to simultaneously fit the scattered light and SED.  A distribution of silicate grains successfully matched the \bFW-band profiles and SED, and in a separate fit, the \bH-band profiles and SED could be matched using a minimum grain size that is 10 times larger than the visible case.  This discrepancy implies that size distributions which match the scattered light colors underestimate the amount of large grains responsible for the sub-mm emission.
A different approach was taken by~\citetalias{strubbe&chiang06}, who modeled the dynamical structure of the debris disk to match the \bFW\ scattered light and SED.  For sufficiently large values of the stellar mass loss rate $\dot{M}$, this model produces a blue color, with an outward blue gradient.

These previous efforts highlight two essential results.  First, the blue color of the scattered light suggests small dust grains scattering in the Rayleigh regime, i.e., submicron particles.  Seceond, the shallow slope of the long-wavelength SED is best fit with large grains of up to mm size (grains larger than this are weakly constrained).
More recently,~\citet{graham_etal07} demonstrated that the linear polarization of the \bFW\ scattered light requires porous grains.  These data are important because the degeneracy between scattering asymmetry and spatial distribution can weaken inferences of debris disk structure based solely on measurements of total intensity.  As the structure, size, and compositions of the grains determine the scattering properties, the assumption of compact grain types in previous work warrants reexamination.

Our goal is to find the simplest description of the grains and their distributions that is compatible with the variety of available observational data.
In \S\ref{subsec:dustmodeling} we describe the methods and structure of our models, the process used to fit the models to the observed scattered light and SED, and the properties of the resulting best-fit models.  We then examine these results in the context of the dynamical model of~\citetalias{strubbe&chiang06} for a steady-state grain distribution produced by a ring of parent bodies (\S\ref{subsec:birthring}).

\subsection{Dust Modeling}\label{subsec:dustmodeling}

\subsubsection{Monte Carlo Radiative Transfer}\label{subsubsec:montecarlo_radtrans}

We model the debris disk around \AUMic\ using \MCFOST, a Monte Carlo radiative transfer code that uses the Stokes formalism to treat the interaction between dust grains and photons to produce SEDs, scattered light images, and polarization maps.  \MCFOST\ is fully described in~\citet{pinte_etal06}, and we summarize its main features here.  Monochromatic photon packets are emitted by the star and propagated through the disk.  The optical depth through which the photon travels before it interacts with a dust grain is randomly chosen from an $e^{-\tau}$ probability distribution.  Scattered light and the thermal SED are computed separately.  When monochromatic scattered light images and polarization maps are produced, photons are only allowed to scatter off dust grains, with a loss of intensity corresponding to the absorption cross-section of the grain. When the SED is computed, photons are either scattered or absorbed, depending on the local albedo; in the latter case, they are immediately re-emitted at a longer wavelength selected on the basis of the local dust temperature.  Once photons exit the computing volume, they are stored in ``reception captors,'' corresponding to specific inclinations.  Maps at all inclinations are simultaneously created, but in this study we focused on the most edge-on captor, which includes inclinations ranging from 88\fdg9 to 90\degr, believed to be appropriate for the inner disk of \AUMic.

To calculate the thermal equilibrium of dust grains, and therefore SEDs, the dust properties must be known throughout the electromagnetic spectrum, from the ultraviolet to the millimeter regime.  It is not possible to describe the dust grains with a simple parametrization based on the albedo and the phase function asymmetry factor, for instance, unless these are known at all wavelengths.  Rather, \MCFOST\ relies on Mie theory (i.e., the grains are assumed to be spherical or essentially randomly oriented), so the knowledge of the dust optical indices (the complex index of refraction) at all wavelengths is sufficient.  Grains with complicated structure, such as porous aggregates, are approximated by spheres with an effective optical index at each wavelength.  Mie theory is used to calculate the scattering and absorption properties for these effective-medium spheres.  In the case of \AUMic, small Rayleigh-scattering grains are indicated by the blue color and high polarization fraction~\citep{graham_etal07}.  For composite grains, an effective medium approximation (depending on the constituent materials and method for computing the effective indices) can be reasonably accurate in the Rayleigh limit~\citep{voshchinnikov&mathis99}.

\MCFOST\ was first developed to model gas-rich, optically thick disks surrounding T\,Tauri stars.  Several features have been added to efficiently and correctly treat the case of optically thin debris disks like \AUMic.  The first modification was to increase the computational efficiency of treating optically thin material.  Rather than expend effort computing the fates of all random photons, the majority of which will not scatter off grains in the optically thin disk, we enforce the first scattering of each photon packet to occur within the disk.  The expected non-interacting photons are not randomly generated; instead we analytically account for the corresponding transmitted starlight.  This ensures that all randomly generated stellar photons scatter at least once in the disk without energy loss.  Second, in the absence of gas and given the low density of dust particles, the dust grains cannot be considered to be thermally equilibrated with one another.  Rather, we compute a size-dependent temperature for the grains, each being in equilibrium with the surrounding radiation field.  In the case of the \AUMic\ disk, this results in a significantly different shape of the SED in the mid-infrared due to large difference in temperature between grains of different size.  Future high-resolution spectra in the 10\,\micron\ region may measure features which can constrain the composition and size distribution of grains~\citep[e.g.][]{li&greenberg98, chen_etal06}.

\subsubsection{Model Construction}\label{subsubsec:model_constr}

Disk models frequently assume power-law descriptions for both the geometry of the disk and the grain size distribution.  This is unlikely to reproduce all observed aspects of the disk, such as the complex small-scale structure seen in the scattered light (\S\ref{subsec:structure}).  It is nonetheless a valuable approach for constraining some of the main system parameters and testing simple hypotheses.
Here, we attempt to reproduce the observations of the disk (SED and scattered light profiles) with a two-zone disk description.  In a narrow inner zone, large grains would be present and account for most of the long wavelength thermal emission, whereas a much more extended outer region would contain small grains and be responsible for the scattered light.  This is a model that qualitatively matches models proposed by~\citetalias{strubbe&chiang06} and~\citet{augereau&beust06}, where our inner annulus would represent the observable population of parent bodies and the outer zone the populations of small grains created by collisions of large bodies and swept out by pressure forces.

In both regions, we assume that the surface density and the grain size distributions follow $\Sigma(r) \propto r^p$ and $dN(a) \propto a^\gamma da$.  The outer radius is fixed at 300\,AU, as scattered light has been detected out to a sensitivity limit of 210\,AU~\citep{kalas_etal04}.  The inner radius and transition radius between the inner and outer regions are free parameters.  In addition, the surface density index, minimum and maximum grain size and distribution index, and the total dust mass in each of the two zones are free.  We initially use a $\gamma=-7/2$ index for the size distribution, which is commonly assumed for debris systems and is suitable for a steady-state collisional cascade~\citep{dohnanyi69}.
While both theory and observations may suggest specific functional forms for the vertical density profile of the disk, we cede to our preference for simplicity and assume that the vertical density distribution follows a Gaussian with a fixed width of $\sigma=0.8$\,AU and a flat radial dependence.  In this case, the photometric apertures capture the majority of the scattered light flux during brightness profile calculation.  The particular choice of vertical density function is unimportant as long as the scale is much less than the photometric aperture height~(\S\ref{subsec:sb_prof}).  We defer for future modeling the exploration of relationships between the form of the vertical density profile, the dependence of scale height with radius, grain scattering phase functions, and the geometry of the disk.

The data for the SED were first compiled by~\citetalias{metchev_etal05} and supplemented with new \textit{Spitzer} measurements by~\citet{chen_etal05}.  We use a $T_\mathrm{eff}=3600$\,K, $\log g=4.5$ \NextGen\ model for the stellar photosphere~\citep{hauschildt_etal99}, using radius $R=0.88 \,R_\sun$ to match the observed stellar flux.  For scattered light, we use the \bJ-, \bH-, and \bKp-band profiles from data presented here, as well as the \bFW-band profiles from data initially presented by~\citetalias{krist_etal05} and re-analyzed for this work in~\S\ref{subsec:sb_prof}.

\subsubsection{Fitting Procedure}\label{subsubsec:fitproc}

Thorough exploration of the available parameter space is a time-consuming task when producing a SED and multiple scattered light images.  We therefore explored it by hand, iteratively narrowing down the possible values for each parameter.  Our objective is not to find the best possible model, but to determine if there is at least one solution based on this simple two-zone model that can match the data relatively well, which would in turn provide support to theoretical birth ring models~(\citetalias{strubbe&chiang06};~\citealt{augereau&beust06}).  We compare our results to these models in~\S\ref{subsec:birthring}.  To further reduce computation, we have elected to fit the \bFW- and \bH-band profiles and the SED, and to later check for consistency with \bJ- and \bKp-band profiles and \bFW\ fractional polarization curves of~\citet{graham_etal07}.  We calculate the model profiles with the same rectangular aperture sizes as the observed profiles in~\S\ref{subsec:sb_prof} (0\farcs 1$\,\times\,$0\farcs 5).

We used a grain model that intrinsically includes a high porosity, a characteristic that provides both strong forward scattering and polarization --- necessary features as demonstrated by~\citet{graham_etal07}.  The dust model developed by~\citet{mathis&whiffen89} consists of a mixture of silicate, carbonaceous, and icy small elements combined into porous aggregates.  With $a_\mathrm{min}=5$\,nm, $a_\mathrm{max}=0.9$\,\micron, a size distribution index of $\gamma=-3.9$, and a vacuum fraction of 80\%, this model reproduces the interstellar extinction law from the ultraviolet to the near-infrared.  We use the same optical constants but allow the size distribution to vary.  We fixed the inner zone's size-distribution index to $\gamma=-3.5$.  We searched for the power law distributions that would adequately produce the blue color (in terms of albedo\,$\times$\,opacity) and roughly wavelength-independent phase function, so as to maintain roughly parallel surface brightness profiles from \bFW\ to \bH.  It was then possible to adjust the model parameters until a match to the SED and the scattered light profiles was obtained.
We tuned the boundaries, grain size distributions, surface density indices, and total mass of each zone.
For comparison, we also fit a model using the optical properties of compact silicate grains (from~\citealt*{draine&li01}) using the same procedure.

\subsubsection{Fit Results}\label{subsubsec:fitresults}

The parameters of our best-fit models are listed in Tables~\ref{tab:modelparms} and~\ref{tab:modelscatparms}, and the results are shown in Figures~\ref{fig:model} and~\ref{fig:model_JK_pol}.
Considering the \bFW- and \bH-band profiles and the SED, we obtain reasonable agreement with the predictions our two-zone model using porous grains: a region of large grains between 35--40\,AU, and a region of smaller grains outward of 40\,AU.
These are slightly further out than the best-fit regions using compact grains, which are more isotropically scattering.
We find that, if the particles in the inner region are porous, they must be in the mm-regime (sizes of 3--6\,mm).  Allowing the size distribution to encompass smaller grains results in an overly steep long-wavelength SED as well as a flux deficiency in the outer small grain region.  Much larger bodies could be present yet undetectable; they would emit very little, even at mm wavelengths.
For porous dust in the outer region, we require a distribution encompassing small grains (0.05--3.0\,\micron, $dn\propto a^{-4.1}da$), with a power-law index that is steeper than the collisional steady-state value of -3.5.

The masses in the inner and outer regions of the porous grain model are around $1.0\times 10^{-2}$\,$M_\earth$ and $2.3\times 10^{-4}$\,$M_\earth$, respectively.
The surface density profile falls rapidly outside of the transition radius (index -2.5 to -3.0), as expected from swept up material; shallower indices yield an improved SED but insufficient scattered flux inside 40\,AU and too-shallow profiles overall.  Steeper indices yield an excessively precipitous \bFW-band brightness profile and excess flux around 50--100\,\micron.  In the inner region, the surface density can be flat or increase with radius, peaking at the transition radius.  This is only loosely constrained: the big grains never get very hot in this narrow annulus and there is little effect on the SED.

We performed some exploration of the parameters to estimate uncertainties.
The masses are relatively well constrained, to 50\% or so, depending on the grain size distribution parameters.  Again, more mass could be hidden in the form of very large bodies in the inner region; the mass derived here is just what is needed to reproduce the long wavelength end of the SED.
In computing the total intensity of scattered light, there is an anticorrelation between $a_\mathrm{min}$ and the region boundary at $\sim 40$\,AU.  We can adjust the minimum grain size by a factor of 2, and the ring radius by $\pm 5$\,AU and still achieve satisfactory \bFW- and \bH-band profiles and SED.

As demonstrated by~\citet{graham_etal07}, the fractional linear polarization profiles provide a complementary constraint on the optical properties of the small grains.  The measured fractional polarization gradually rises to a plateau of $\sim$\,35\% at around 50\,AU.  Our best-fit models using compact and porous grains are comparable in fit quality for the total scattered light intensity profiles and the SED; however the polarization curves in Fig.~\ref{fig:model_JK_pol} clearly rule out compact grains, which reach maximum fractional polarizations $<10$\%.  The porous grains generally follow the measured trend, and reach a plateau of 30--35\% by 40\,AU, the starting location of the small outer grain region.

This model has some shortcomings, in that it predicts excess flux around 60--70\,\micron\ and a slightly too-shallow scattered-light profile in the \bH\ band outside of the transition radius.  This latter point results in a predicted disk color that does not change outside of 1\arcsec, as opposed to the observations.
We also note that the model profiles show less agreement with the \bJ- and \bKp-band profiles (Fig.~\ref{fig:model_JK_pol}).  The overall flux levels differ (though this may coincide with photometric calibration uncertainty; cf.~\S\ref{subsec:calibration}), and we overpredict the midplane surface brightness in the outer regions.
For this analysis, we have avoided goodness-of-fit metrics like $\chi^2_\nu$ because our model, by design, will fail to reproduce substructure in the scattered-light profiles.  Further, PSF subtraction residuals introduce correlations between errors in the measured profiles, which we ignore (\S\ref{subsec:sb_prof}).  Nevertheless, we can get a sense of model fidelity by considering the long wavelength end of the SED (where the dust contributes).  For the nine SED points in Fig.~\ref{fig:model} at $\lambda>10$\,\micron, the best-fit compact grain model gives $\chi = 8.6$ and the porous model gives $\chi = 10.9$.  The bulk of the deviation arises from the 60--70\,\micron\ region.  A formal $\chi^2_\nu$ analysis is not applicable, as we have also tuned our parameters to match the scattered light.
In contrast to the porous grain model, the compact grain model underpredicts the 60--70\,\micron\ flux.  With their higher average albedo, the compact grains absorb less energy than the porous ones (Table~\ref{tab:modelscatparms}), suggesting a better fit may be obtained with grain albedos in the 0.6--0.7 range.
Still, the fit is satisfying, considering the simplicity of the model~(\S\ref{subsubsec:model_constr}).  Presumably, considering non-power law prescriptions, other dust composition, non-spherical grains, and/or overlap between small and large grains would result in a better fit.  One attractive possibility, that we do not explore here, is that ever-smaller grains are present as we move outside of the belt of parent bodies~(\S\ref{subsec:birthring}).  This could account for the increasingly blue color in the outer region as well as for the seemingly different transition radii obtained from the \bFW- and \bH-band profiles ($\sim$\,40\,AU vs. $\sim$\,30\,AU).

\subsection{Birth Ring Examination}\label{subsec:birthring}

In the Solar System, the Asteroid Belt and Classical Kuiper Belt constitute rings of solid bodies that, through mutual collisions, act as sites of continual dust production.  Several forces act on the newly liberated grains, and the resulting trajectories shape the overall structure of the Sun's debris disk.  
A similar belt of parent bodies has been invoked to explain the break in the surface brightness profile of the \betaPic\ debris disk~\citep{lecavelier_etal96,augereau_etal01}.  Additional rings have been proposed to explain features seen in recent imaging~\citep[e.g.][]{wahhaj_etal03,telesco_etal05} and spatially resolved spectroscopy~\citep{weinberger_etal03,okamoto_etal04} of this system.
An attractive feature of such models is that their dust distributions are in steady-state --- we need not appeal to a rare (but recent) catastrophic collision between large (e.g. km-sized) bodies, such that fresh grains have had insufficient time to completely diffuse through the system.

The breaks in the slopes of \AUMic's surface brightness profiles naturally raise the prospect of a ring of parent bodies analogous to the Classical Kuiper Belt.  The existence of such a belt was proposed by~\citetalias{metchev_etal05},~\citetalias{strubbe&chiang06}, and~\citet{augereau&beust06}.  Here, we focus our attention on a detailed theoretical model, developed by~\citetalias{strubbe&chiang06}, which yields the steady-state spatial and size distributions of dust grains in the disk.  
A critical parameter is the stellar mass-loss rate, $\dot{M}_*$, which governs the corpuscular forces on the grains.  In addition to modifying the disk structure, the stellar wind is crucial in determining the lifetimes of debris disks around late-type stars~\citep{plavchan_etal05}.  Assuming the scenario of steady-state dust production by a ring of parent bodies is applicable, one may use a dynamical model to infer $\dot{M}_*$ from observations.
We note that the theory was applied to the \AUMic\ disk prior to the observational findings of~\citet{graham_etal07}, who showed that the disk's scattered light is strongly polarized, and that porous grains are required to fit the polarization profile.  In this \doctype, we examine the disk structure predicted by the birth ring scenario in light of the currently available observational data.  After a brief review of the predictions for compact grains, we probe modifications to the theory given the porous grains of our best-fit model and check the consistency of the theory's features with those of our Monte Carlo radiative transfer model.  In this context, we appraise the predictive power of the observations to determine $\dot{M}_*$ via inferences of the size and space distributions of grains.

\subsubsection{The Theoretical Scenario}\label{subsubsec:theory}

We first recapitulate the basic physical arguments of the model put forth by~\citetalias{strubbe&chiang06}.
Collisions occur within the ring of parent bodies, producing dust grains.
The radial components of forces arising from stellar radiation and wind result in a pressure acting to push grains away from the star, causing blow-out for sufficiently small grains.
In contrast, the tangential components of these forces act as a drag mechanism, which for radiative forces is the well-known Poynting-Robertson (PR) drag.  Grains dominated by drag forces spiral in toward the star.
The competition between these forces and the relative frequency of destructive grain collisions determine the dynamical structure of the disk.  In disks dominated by drag forces, (\citetalias{strubbe&chiang06}'s ``type A''), the action of PR and corpuscular drags fill the region interior to the birth ring with grains that avoid destructive collisions.  In disks dominated by collisions (``type B''), this interior is essentially empty, as the pulverized remains of colliding grains are quickly swept outward.  In both cases, the region exterior to the birth ring is largely populated by the small, tenuously bound grains which follow elliptical orbits with periastra near the birth ring.  In type B disks, these grains are just larger than the blow-out size of stellar wind and radiation pressure.  In type A disks, the smallest grains are pulled inward by corpuscular and PR drag, and the peak of the grain size distribution in the outer disk corresponds to a size significantly larger than that of blow-out.

\citetalias{strubbe&chiang06} developed analytical and numerical descriptions of the results of these processes on the size-space distributions of dust grains, and then applied this theoretical framework to the case of \AUMic.  They postulate a belt of parent bodies at 43\,AU, a location determined by the surface brightness profile break in \textit{HST} imaging.
By simultaneously modeling the scattered light profiles (\bFW-band from~\citetalias{krist_etal05} and \bH-band from~\citetalias{metchev_etal05}) and the SED with compact spheres of pure water ice, they determine that type B conditions hold and place limits on $\dot{M}_*$.  With these assumptions on grain type, there are two pieces of evidence for an inner disk devoid of in-spiraling grains, which is expected for type B conditions: (1) a lack of photospheric excess in the 10\,\micron\ region of the SED, and (2) the shallow slope of the surface brightness profile at projected distances interior to the proposed ring.
\citet{augereau&beust06} argue for a similar dynamical scenario for explaining the observed structure, where corpuscular pressure forces sweep grains outward.  Considering the pressure from the stellar wind (and additional radiative enhancement by stellar flares), they determine that radial forces can sufficiently diffuse small grains into the outer regions to account for the observed profile break, in analogy with the dominance of radiation pressure in \betaPic's disk.

\subsubsection{Potential Ramifications of Porous Grains}\label{subsubsec:porous_comp}

It is important to examine the birth ring scenario in view of the new constraints on scattering properties from recent imaging of the disk in polarized light~\citep{graham_etal07}.  The polarization data strongly indicate the presence of porous grains scattering in the Rayleigh regime, reaffirmed in our modeling~(\S\ref{subsubsec:fitresults}).  Such grains are largely forward-scattering and produce the required high peak polarization fraction.  
Since scattering asymmetry directly affects the inferred spatial distribution of grains, previous models' use of spheres of solid material must be re-evaluated with this new evidence.
We also stress that, as noted in~\S\ref{subsec:sb_prof}, analyses involving surface brightness profiles require consistent methodology across all data sets.
Following these points, it is appropriate to both check the consistency of the model by~\citetalias{strubbe&chiang06} with our porous grain Monte Carlo model, and to evaluate the strength of such models for inference of \AUMic's stellar mass-loss rate.

In approaching this problem, it is apt to contrast the relevant properties of the compact and porous grain types we use in the models of~\S\ref{subsec:dustmodeling}.
As expected, the porous grains of~\citet{mathis&whiffen89} are less dense than the compact grains, at $\rho=0.5$\,g\,cm$^{-3}$ (compared to 2\,g\,cm$^{-3}$ for~\citealt{draine&li01} grains).
At a given size, the porous grains also couple less efficiently to the radiation field than their compact counterparts.  Taking a 0.05\,\micron\ grain as an example, the effective cross section for radiation pressure ($Q_\mathrm{pr}\pi a^2$) is more than an order of magnitude smaller for a porous grain: $Q_\mathrm{pr}\sim 2\rightarrow 0.05$ when changing from compact to porous.
This coupling inefficiency is similarly manifested in a decrease in the effective cross section to scattering, $Q_\mathrm{sca}\pi a^2$, for porous grains.  

Debris disk structure is commonly parameterized by the vertical optical depth.  In the following analysis, we will refer to both the vertical optical depth to scattering, \tausca, and the geometric vertical optical depth, \taugeo.  The former quantity is useful for describing the 1-d brightness profile, while the latter governs the collisional timescale.  For a given surface mass density $\Sigma$, the depth \taugeo\ depends only on the grain density and size distribution $dN/da$, while \tausca\ additionally depends on scattering efficiency:
\begin{eqnarray}
\taugeo(r) &\propto& \frac{\Sigma(r)}{\rho}\int \left(\pi a^2\right)\frac{dN}{da}da, \label{eq:taugeo}\\
\tausca(r) &\propto& \frac{\Sigma(r)}{\rho}\int \left[Q_\mathrm{sca}(a)\pi a^2\right]\frac{dN}{da}da. \label{eq:tausca}
\end{eqnarray}
For a given grain type and size distribution, \taugeo\ and \tausca\ can differ by orders of magnitude.
As an example, consider the outer region of porous grains in our model from the previous section.  While the size distribution is steep ($dN/da\propto a^{-4.1}$), because $Q_\mathrm{sca}(1\,\micron)\gg Q_\mathrm{sca}(0.05\,\micron)$, micron-sized grains contstitute the bulk of $\tau_\mathrm{sca}$ despite their relative scarcity.  The geometric depth $\tau_\mathrm{geo}$, on the other hand, is dominated by the smallest grains, resulting in $\tau_\mathrm{geo}\gg\tau_\mathrm{sca}$.

\subsubsection{Disk Structure Analysis: Theoretical Considerations}\label{subsubsec:structure:analytical}

The models of~\citetalias{strubbe&chiang06} delineate three regions of water ice grains: (1) a birth ring, populated by parent bodies of sizes up to $\sim$10\,cm, (2) an outer region made up of smaller grains ($a_\mathrm{min}\lesssim 1$\,\micron) on loosely bound orbits, and in type A disks, (3) an inner region of grains, $\lesssim$\,\micron\ size, spiraling inward as a result of corpuscular and PR drag forces.  Grains smaller than the stellar wind and radiation pressure blow-out size, $a_\mathrm{min}$ ($\sim$\,0.1\,\micron), have negligible contribution to the disk.  Our modeling of the dust distribution with porous grains~(\S\ref{subsec:dustmodeling}) shows that a two-zone architecture (lacking an inner region of drag-dominated grains) is consistent with the observations and, qualitatively, with the architecture predicted by physically modeling dust generated by a belt of parent bodies.

In the birth ring model, how does the expected grain size distribution change in light of the differences between the~\citet{mathis&whiffen89} grains and water ice spheres?
We first consider $a_\mathrm{blow}$, which is determined by the combined action of radiation and stellar wind pressures.  The density decrease of porous grains relative to spheres tends to increase $a_\mathrm{blow}$.  However, this effect is offset by the decrease in the effective cross section, parameterized by $Q_\mathrm{pr}$ (a factor for 40 lower for $a=0.05$\,\micron\ grains), resulting in a net decrease of porous grains' blow-out size relative to their compact counterparts.  Rewriting~\citetalias{strubbe&chiang06}'s Eqs.~7 and~8, the blow-out size assuming constant pressure is
\begin{equation}
a_\mathrm{blow} = \frac{3}{8\pi}\frac{L_*}{GM_*c\rho}
\left(Q_\mathrm{pr}+Q_\mathrm{wind}\frac{\dot{M}_*v_\mathrm{wind}c}{L_*}\right).
\end{equation}
We follow~\citetalias{strubbe&chiang06} in adopting $v_\mathrm{wind}=450$\,km s$^{-1}$ and $\dot{M}_\sun=2\times 10^{-14} M_\sun$ yr$^{-1}$.  We choose $L=0.12$\,$L_\sun$ and assume that the cross section to wind ($Q_\mathrm{wind}\pi a^2$) is unchanged ($Q_\mathrm{wind}=1$) for the porous grains.  Together these yield
$a_\mathrm{blow} = \left\lbrace\right.$0.029, 0.047, 0.23, 2.1$\left.\right\rbrace$\,\micron\
for
$\dot{M}_* = \left\lbrace\right.$1, 10, $10^2$, $10^3\left.\right\rbrace \dot{M}_\sun$.    The minimum grain size in the outer region of our disk model (0.05\,\micron) is larger than $a_\mathrm{blow}$ for the weaker stellar winds, which suggests $\dot{M}_*\ll 10^2\dot{M}_\sun$.

Next, we consider the implications of our porous grain model on the amount of disk infill (type A vs. B) predicted by the birth ring theory --- a process regulated by the collision rate, which depends on  \taugeo\ (Eq.~\ref{eq:taugeo}).
In our model for the scattered light at $r_\mathrm{BR}=40$\,AU, the smallest grains contribute the bulk of the total geometric cross section.  The abundance of these grains increases \taugeo\ at the birth ring relative to the~\citetalias{strubbe&chiang06} value by a factor of 10, to $\taugeo(r_\mathrm{BR})=4\times 10^{-2}$.  This shortens the timescale for destructive collisions of the average particle, assuming the specific collisional energy is unchanged.
The collisional timescale is balanced by the drag-induced infall time at a grain size defined by $a_\mathrm{break}$ (cf.~\citetalias{strubbe&chiang06} Eqs.~36--37).  Grains near this size constitute the bulk of $\taugeo(r_\mathrm{BR})$, and in equilibrium, grains of sizes $a_\mathrm{blow}<a<a_\mathrm{break}$ fill the interior ($r<r_\mathrm{BR}$).  Like the case of compact ice spheres, for porous grains we find that $a_\mathrm{break}$ is the same order as $a_\mathrm{blow}$: $a_\mathrm{break}=\left\lbrace\right.$1.005, 1.03, 1.08, 1.1$\left.\right\rbrace a_\mathrm{blow}$ for a peak geometric vertical optical depth at $r_\mathrm{BR}$ and the same stellar wind values.
In contrast to~\citetalias{strubbe&chiang06}'s results for the compact grain case, we find that $(a_\mathrm{break}-a_\mathrm{blow}) < a_\mathrm{blow}$ for porous grains in all wind cases, including stronger values.  The implication of this result is that no amount of corpuscular drag will produce a significant fraction of grains with orbits crossing the inner region, favoring a clear inner disk (type B).

\subsubsection{Disk Structure Analysis: Observational Limits on Infill}\label{subsubsec:structure:observational}

As a complementary approach to examining the birth ring scenario, we can, for the moment, set aside the final conclusion of the previous section and use our model from \S\ref{subsec:dustmodeling} to test allowed small-grain infill.  When using \MCFOST\ to model the observed disk with porous grains, can we make a distinction between the conditions of type A and B?
By definition, our two-zone model lacks grains inward of the putative birth ring, consistent with a collisionally dominated type B disk.  To test whether the system could be consistent with type A, we now consider adding a third region of grains.
If a significant mass of small grains are present inward of $\sim$\,35\,AU (type A), then these warmer grains would increase dust emission in the 10--20\,\micron\ region of the SED and add to the scattered light of the system (Figs.~\ref{fig:model}--\ref{fig:model_JK_pol}).
We note that, compared to more isotropically scattering spheres, the strong forward scattering of porous grains allows us to hide more mass in this region without strongly affecting the total intensity of scattered light.  The smallest projected separations, where the added contribution from forward scattering would be maximal, are lost in the glare of the central star and are behind the occulting spot.

The additional zone of porous grains has a flat $\Sigma(r)\propto r^0$ density profile and ranges from 15--35\,AU.
The inner cutoff corresponds to the estimate of the ice boundary by~\citetalias{metchev_etal05} (13--15\,AU).
As a general feature of the birth ring conditions modeled by~\citetalias{strubbe&chiang06}, the outer region's grains have $a_\mathrm{min}\sim a_\mathrm{break}\sim a_\mathrm{blow}$; therefore we model a single-sized population of $a=0.05$\,\micron\ grains in this new inner region.
The middle and outer regions' dust distributions were fixed to the best-fit parameters (Table~\ref{tab:modelparms}), and we adjusted the mass in the new inner region in $\sim $\,0.5 dex steps until the scattered light profiles (including fractional polarization) or SED showed significant deviation.
We found that the SED first showed noticeable deviation at $M=10^{-5} M_\earth$; this is an upper limit to the mass of grains in this population allowed by the observations.

We can express this limit as either the geometric or scattering vertical optical depths relative to the peak depth in the birth ring, $\zeta\equiv\tau_\perp(r<r_\mathrm{BR})/\tau_\perp(r_\mathrm{BR})$.  The geometric vertical optical depth is more closely related to the dynamical structure, while the scattering depth is traced by our imaging observables.  As noted in~\S\ref{subsubsec:porous_comp}, these geometric and scattering optical depths can be quite different.
We find $\taugeo(r<r_\mathrm{BR})\lesssim 2.5\times 10^{-2}$, giving $\zeta_\mathrm{geo}\lesssim 0.6$.
While our limit to $\zeta_\mathrm{geo}$ is order unity, the strong decline in $Q_\mathrm{sca}$ with grain size results in a more stringent limit to $\zeta_\mathrm{sca}$ for 0.05\,\micron\ grain filling.  An inner mass of $3\times 10^{-5} M_\earth$, easily ruled out by our analysis, corresponds to $\zeta_\mathrm{sca} \sim 1/400$.  This is the same order as the limit of~\citet{graham_etal07}, who determined $\zeta_\mathrm{sca} \lesssim 1/300$ inward of 40\,AU using a similar method (though without SED modeling).

While the limit for $\zeta_\mathrm{sca}$ implies an inner disk relatively free of scattering grains, the relevant quantity for the dynamical structure of the disk is $\zeta_\mathrm{geo}$.  The weakness of our geometric optical depth ratio limit implies that we cannot yet determine type A vs. B dynamics or make inferences of $\dot{M}_*$ by matching the expected disk structure to current scattered light and SED data.  Rather, perhaps the strongest direct observational evidence for a type A disk would come from resolved thermal emission from this zone.

\subsubsection{Future Considerations}\label{subsubsec:future}

When considering more rigorous tests of birth ring theory, several enhancements can be made to our simple disk model.
Faithful models must not rely on an artificial separation between the site of dust production (inner, large-grain region) and the small grains, as we have done in~\S\ref{subsec:dustmodeling}.  
\citetalias{strubbe&chiang06} also predict different asymptotic power-laws for the outer region's geometric vertical optical depth in type A and B disks (-2.5 and -1.5, respectively).  However, the corresponding asymptotes in scattered light profiles are not reached until distances of a few 100\,AU, ruling out reliable comparisons with current data.
Somewhat more challenging aspects of the theory are grain size distributions that do not follow power-law description (cf. Fig.~3 of~\citetalias{strubbe&chiang06}).
The prediction of~\citetalias{strubbe&chiang06} is an excess of grains above the Dohnanyi collisional cascade ($dn\propto a^{-7/2}da$) for sizes near $a_\mathrm{blow}$.  Incorporating a population of grains at this size, in addition to the power-law size distribution, is a possible avenue for future significance testing.
Both~\citetalias{strubbe&chiang06} and~\citet{augereau&beust06} predict diminishing minimum grain size with increasing stellar distance (due to ever-more tenuously bound grains as a result of stellar wind and radiation pressure), which can effectively model the observed blue color gradient in the outer region.  Future scattered light modeling may incorporate a function $a_\mathrm{min}(r)$ in the outer zone.

\section{DISCUSSION \& CONCLUSIONS}\label{sec:discn&concl}

Debris disks are a long-lived phase in the evolution of circumstellar material, and they provide indirect probes for the presence of planets and planetesimals.  Resolved imaging is crucial for determining the structure and dynamical history of the dust.
In our near-IR imaging, we measured the structure of the disk seen in scattered light.  We confirmed the overall architecture of the disk seen in \bH\ band by~\citetalias{liu04} and~\citetalias{metchev_etal05}, and presented the first images seen in \bJ\ and \bKp\ (\S\ref{subsec:structure}).
We represented the global disk structure with a power-law description of the midplane surface brightness, and noted that discretion must be used when comparing between profiles measured in different manners (\S\ref{subsec:sb_prof}).
Scattered light features on both large and small scales are largely consistent across the bands studied here.
The colors we measured in the surface brightness profiles (\S\ref{subsec:scat_col}) allow inferences of the scattering properties, which are determined by the composition, sizes, and structures of dust grains.

We used a model for the space and size distributions of the grains to reproduce these profiles and the SED (\S\ref{subsec:dustmodeling}).
The simple, two-zone description of power-law grain distributions cannot faithfully reproduce the observations with compact silicate grains.  Rather, porous grains of dirty ice, whose presence is indicated by recent polarization measurements~\citep{graham_etal07}, have optical constants with which we can match such an architecture to the data with reasonable fidelity.
The key degeneracy is between the scattering phase function and the density distribution.
Models of~\citetalias{krist_etal05} and~\citetalias{metchev_etal05} place grains which exhibit moderate scattering asymmetry in the inner regions of the disk.  The porous grains in the outer zone of our model, which are larger and more forward scattering than the compact grains used in previous efforts, fill in this region of scattered light.  As such, our modeling does not require grains inward of 35\,AU to reproduce the midplane surface brightness profiles.
The inner zone (35--40\,AU) of our model is made up of few-mm-sized grains, which reproduce the far-IR end of the SED.  The outer zone grains of our model are smaller (0.05--3.0\,\micron) and produce the majority of the mid-IR emission and the scattered light.

The blue color of the scattered light disk is relatively rare among resolved systems~\citep[cf. Table~1 of][]{meyer_etal07}.  The succession of modeling efforts have shown that a significant population of submicron grains reproduces the disk's blue scattered-light color, and that the presence of such grains is compatible with interpretations of the disk's dynamical structure.
As other authors have noted, stellar properties play a major role in the removal of submicron grains.  As an M dwarf, \AUMic\ has a uniquely low radiation pressure relative to other stars with scattered-light debris disks.  For sufficiently low values of $\dot{M}_*$, we find $a_\mathrm{blow}$ is submicron and grains will scatter blue.
While stellar properties can reasonably be used as a predictor for the presence of submicron grains in debris systems, grain properties can also affect the blow-out size.  As we noted in~\S\ref{subsubsec:porous_comp}, the smallest of our porous grains have much smaller $Q_\mathrm{pr}$ relative to their compact counterparts of the same size.  The blow-out sizes we calculate for porous grains are up to a factor of eight smaller than those of compact grains (as computed by~\citetalias{strubbe&chiang06}).
When blue scattered-light colors are detected around earlier-type stars with stronger radiation pressure~\citep[e.g. HD\,32297 and HD\,15115;][]{kalas05, kalas_etal07}, inferences of submicron grains may constrain their porosity and composition~\citep[see also][]{artymowicz88, grigorieva_etal07}.

The study of debris disk evolution hinges on accurate measurements of stellar ages and dust masses.
The total observed mass of the disk in our model ($1.0\times 10^{-2}$\,$M_\earth$) is dominated by the mm-sized grains in the inner zone.
Note that this is a factor of few times smaller than the mass we derive for compact grains ($2.7\times 10^{-2}$\,$M_\earth$).  Though the equilibrium temperatures for the two grain types differ, this is primarily an opacity effect~\citep{voschinnikov_etal06}.  The decrease in density increases the opacity, which in turn lowers the mass needed to match the sub-mm fluxes.  Since these grains contribute little to the scattered light, we have no information about their porosity (and correspondingly whether any compaction has occurred).  Therefore, in addition to the normal caveats about dust masses derived from sub-mm fluxes (e.g. insensitivity to larger bodies), derived debris disk masses are additionally uncertain by a factor of a few depending on the grain porosity.  The same opacity caveat applies to the inferred sizes of the grains that dominate the sub-mm emission.  In the inner region, our porous grains are a factor of three larger than the equivalent best-fit compact ones --- the increased opacity means a larger size is needed to match a given temperature.

Nevertheless, we can compare our dust masses to those of other authors.
Our estimate is similar to the mass estimated by a single-temperature fit to the long-wavelength SED by~\citet{liu_etal04}, who found $1.1\times 10^{-2}$\,$M_\earth$ of 40\,K dust at 17\,AU.  \citetalias{metchev_etal05} found a similar mass as~\citeauthor{liu_etal04}, for single-zone power-law distributions of size and density of ISM-like grains.
\citet{augereau&beust06} calculate a mass of $7\times 10^{-3}$\,$M_\earth$ by adjusting the grain size distribution in their scattered light model in order to match the long-wavelength SED.
Finally,~\citetalias{strubbe&chiang06} argue for parent body sizes up to 10\,cm, the maximum size participating in a steady-state collisional cascade.  For their model's size distribution, the mass of parent bodies is $\sim$\,$10^{-2}$\,$M_\earth$.  However, current observations are not sensitive to grains larger than a few mm, so the maximum observed mass in this model is significantly lower.  For size distribution index $\gamma=-7/2$, the mass $M\propto a_\mathrm{max}^{1/2}$, and the observable portion of~\citetalias{strubbe&chiang06}'s mass is a few times smaller than that of our work.

The interpretation of debris disk observations for inferring the presence of planets and the history of the grains rests on sound physical modeling of grain dynamics.
The overall shape of the midplane surface brightness profile distribution is consistent with steady-state dynamics arising from a ring of parent bodies (\S\ref{subsec:birthring}).
With models of grain dynamics, we might constrain $\dot{M}$ by examining the density of small grains in the inner disk ($\lesssim 35$\,AU).  We found that for the porous grains in our model, an inner region would be populated primarily by grains just above the blow-out size for a wide range of $\dot{M}_*$.  Further, the constraint of the presence of grains in an inner region is set by the mid-IR SED, and we can only say that the geometric vertical optical depth in this region is at most the same order as the peak depth in the birth ring, consistent with both drag- and collision-dominated (type A and B) disks.  However, the presence of small (0.05\,\micron) grains in the outer region implies that the mass-loss rate is low enough such that these grains survive blow-out, i.e. $\dot{M}_* \ll 10^2\dot{M}_\sun$.

The overall disk architecture inferred from these models has implications for the substructure seen in scattered light.
In \S\ref{subsec:substructure}, we compared our measurements of features A--E with the results of other authors, giving further evidence that they are not artifacts of stellar PSF subtraction.
Additional brightness enhancements located 11--15\,AU from the star have also been identified, though additional observations of the disk at small angles are needed to improve confidence.
The appearance of such substructural features is determined by the grain distribution; since the bulk of the scattered light comes from small grains beyond $\sim 40$\,AU, the features at projected distances inward of 40\,AU must arise from azimuthal perturbations in the dust distribution beyond this distance.
Some of these features appear to exhibit a wavelength-dependent position.  (We note the position of feature D, which is closest to 40\,AU, does not appear to do so.)
While the overall position of a feature may be used to trace a perturbation in the density distribution in the outer disk, the wavelength dependence of the position may constrain additional spatial variation in the grain size distribution.
In a similar vein, any vertical structure associated with a given feature (Fig.~\ref{fig:vertstructure}) can be tied to the vertical structure of the outer disk.

In summary, we:
\begin{enumerate}
\item Demonstrate a new roll-subtraction technique that attempts to mitigate AO PSF variability~(\S\ref{subsec:psf_sub}, Appendix~\ref{ap:roll_sub}),
\item Detect the \AUMic\ debris disk in \bJ\bH\bKp-band imaging and place an upper limit on the \bLp-band brightness distribution~(\S\ref{subsec:structure}),
\item Confirm the blue color of the disk and measured a blue color gradient outside of the transition radius~(\S\ref{subsec:sb_prof}-\ref{subsec:scat_col}),
\item Place detection limits on point sources in the disk midplane~(\S\ref{subsec:pt_src}) and document a technique for determining point-source detection sensitivity in the disk midplane (Appendix~\ref{ap:pt_src}),
\item Verified the presence of substructure in the inner disk, and shown that some features exhibit slight variation in their positions with wavelength~(\S\ref{subsec:substructure}),
\item Demonstrate the applicability of a simple two-zone model which simultaneously fits the scattered light profiles and SED~(\S\ref{subsec:dustmodeling}),
\item Find that compact silicate grains cannot be used in our simple model, while porous, icy aggregates of silicate and carbonaceous grains can reasonably account for the observed thermal and scattered light~(\S\ref{subsec:dustmodeling}),
\item Determine that a two-zone model is consistent with steady-state grain dynamics dominated by collisions (\citetalias{strubbe&chiang06} type B;~\S\ref{subsec:birthring}),
\item Show that, by relying on models of the inner disk ($r\lesssim 35$\,AU) structure, we cannot place strong limits on the geometric vertical optical depth of small (0.05\,\micron) porous grains interior to the birth ring and therefore cannot yet constrain $\dot{M}_*$ from this approach alone~(\S\ref{subsec:birthring}),
and
\item Show that the blow-out size for porous grains is consistent with modeled grain size distribution for $\dot{M}_*\ll 10^2\dot{M}_\sun$~(\S\ref{subsec:birthring}).
\end{enumerate}

The next modeling steps will seek to utilize all available imaging and SED data.  We note in particular that measurements of scattering in polarized light provide strong, complementary constraints on the composition and distribution of dust in this system.  Future work will incorporate fitting to scattered light images rather than surface brightness profiles, and will begin to probe the vertical structure of the disk while taking into account the effects of projection and blurring by the PSF.
Finally, we emphasize that the global structure of the debris can be explained by steady-state dust production and diffusion, and we need not invoke a planet to clear the inner region of dust.  However, the mechanisms responsible for maintaining the structure of parent bodies in the birth ring as well as the dust substructure remain undetermined.

\acknowledgements
The authors wish to thank John Krist for enlightening the intricacies of \textit{HST} ACS coronagraphic data calibration,
as well as Eugene Chiang, Peter Plavchan, and Linda Strubbe for insightful discussions.
This work was supported in part by the NSF Science and Technology Center for Adaptive Optics, managed by the University of California at Santa Cruz under cooperative agreement AST-9876783.
This work utilized observations with the NASA/ESA \textit{Hubble Space Telescope} obtained at the Space Telescope Science Institute (STScI), which is operated by the Association of Universities for Research in Astronomy.  Support for Proposal number GO-10228 was provided by NASA through a grant from STScI under NASA contract NAS5-26555.  
Some of the data presented herein were obtained at the W. M. Keck Observatory, which is operated as a scientific partnership among the California Institute of Technology, the University of California, and the National Aeronautics and Space Administration. The Observatory was made possible by the generous financial support of the W. M. Keck Foundation.
This publication makes use of data products from the Two Micron All Sky Survey, which is a joint project of the University of Massachusetts and the Infrared Processing and Analysis Center/California Institute of Technology, funded by the National Aeronautics and Space Administration and the National Science Foundation.
{\it Facilities:} \facility{Keck:II (NIRC2)}, \facility{HST (ACS)}

\appendix
\section{ROLL SUBTRACTION FOR AO}\label{ap:roll_sub}

A decisive arbiter of subtraction fidelity is the stability of the stellar PSF.  We refined the roll subtraction process (Fig.~\ref{fig:subtraction_overview}) to simply accommodate the time variability of ground-based PSFs.  In general, the long-exposure AO PSF is a non-trivial function of atmospheric turbulence fluctuations, filtration by closed-loop AO, and additional quasi-static wavefront errors arising from the optics in the system (``static speckles'').   With the PSF structure arising from these effects in mind, we built an empirical model of the PSF for each exposure, and then used these models to optimally subtract the stellar image.
Our proposed deconstruction hinges on the removal of the radial profile from each image.  It is much more challenging to remove the stellar profile from images with face-on disks.  The procedure outlined here is most-suited for imaging edge-on disks from the ground.

\subsection{PSF Model}\label{subsec:psf_model}

The ideal model PSF will reproduce the response of the imaging system to the starlight which has been occulted by the coronagraph.  For our subtraction problem, we are interested in estimating the structure of the stellar PSF's outer regions --- outside of any coronagraphic focal-plane mask.  For the moment, we ignore the focal-plane mask and consider only the case of a monochromatic image of an on-axis star.  We approximate the adaptively corrected long-exposure point spread function $S(\x)$ as the convolution ($\star$) between a ``static'' PSF and a blurring kernel encompassing the time-variable wavefront errors~\citep[e.g.][]{veran_etal97},
\begin{equation}\label{eq:psf_convol}
S(\x) \approx S_\mathrm{s}(\x) \star \Lambda(\x).
\end{equation}
Here, $\Lambda(\x)$ is determined by the variable wavefront errors of partially corrected atmospheric turbulence which, depending on correction level, has a characteristic width $\lesssim\lambda/r_0$ (where $r_0$ is the Fried length).  $S_\mathrm{s}(\x)$ is the PSF that would arise solely from time-independent wavefront errors.  For a circular pupil of diameter $D$ with no obscurations or wavefront errors, $S_\mathrm{s}(\x)$ is the diffraction-limited Airy pattern of characteristic width $\sim\lambda/D$.

With good AO correction, the structure of the long-exposure PSF is characterized by a diffraction-limited core and a seeing-disk halo.  Features like diffraction spikes and individual speckles also play an important role in the structure.  To further motivate our model, we first decompose the static PSF into the sum of its Airy core ($\theta<1.22 \lambda/D$) and all other static speckles ($\theta>1.22 \lambda/D$),
\begin{equation}
S_\mathrm{s}(\x) = S_\mathrm{s,c}(\x) + S_\mathrm{s,sp}(\x).
\end{equation}
Note that since the core is sharply peaked relative to the halo scale in the blurring kernel ($\lambda/D\ll \lambda/r_0$), $S_\mathrm{s,c}(\x)\star\Lambda(\x)$ is approximately proportional to $\Lambda$ itself; the scale factor $\alpha$ is related to the Strehl ratio.  For computational simplicity, we treat the blurring effect of $\Lambda(\x)$ on $S_\mathrm{s,sp}(\x)$ by a scalar factor $\beta$, which also correlates with correction level --- decreased blurring increases peak speckle intensity.  Further approximating~(\ref{eq:psf_convol}), we have
\begin{equation}\label{eq:psf_sum}
S(\x) \sim \alpha\Lambda(\x) + \beta S_\mathrm{s,sp}(\x).
\end{equation}

We have simplified the PSF from a convolution of static and atmospheric terms (Eq.~\ref{eq:psf_convol}) to a linear combination (Eq.~\ref{eq:psf_sum}), and we now seek to increase computational efficiency further.  First, we note that with this linear combination, it is straightforward to generalize to wide-band imaging since $\Lambda$ and $S_\mathrm{s,sp}$ can each be represented by a linear combination of quasi-monochromatic PSFs.  Second, although not strictly true under real conditions, we further assume that the blurring is symmetric in azimuth ($\phi$).  This leads us to isolate the azimuthally symmetric radial profile of each term in equation~(\ref{eq:psf_sum}).  For an exposure $i$, we estimate the image of the star as
\begin{equation}\label{eq:psf_model}
\hat{S}_i(\theta,\phi) \equiv \rho_i(\theta) + \beta_i S'_\mathrm{s,sp}(\theta,\phi),
\end{equation}
with $S'_\mathrm{s,sp}(\x)$ given by $S_\mathrm{s,sp}(\x)$ with the radial profile removed and subsumed into the profile term $\rho(\x)$.

We treat the coronagraphic focal mask in our model (Eq.~\ref{eq:psf_model}) simply.  \textit{Prima facie}, the circular mask's suppression of the stellar image will be absorbed into $\rho$.  Errors in centering the star behind the mask will result in additional leakage of light outside the spot via diffraction~\citep{lloyd&sivaramakrishnan05}, an effect we ignore.  These misalignments manifest themselves as subtraction errors in the vicinity of the mask edge.

A key requirement for our model is the profile-removed speckle map, $S'_\mathrm{s,sp}(\x)$.  In practice, we use the target images themselves to construct $S'_\mathrm{s,sp}(\x)$.  We remove the radial profile from each image, mask out the region around the (rotating) edge-on disk, and combine the results to form the speckle map.  We additionally mask the the diffraction spikes from each image when estimating the radial profiles $\rho_i$, because the spikes' blurring (which we treat only by a scale factor $\beta$) can bias the profile estimates.  In practice, we mask the disk with a 0\farcs 5-wide strip, and the spikes with regions that are 1\farcs 6 wide, tapering inward for $\theta\leq5\arcsec$.  With $\rho_i(\x)$ and $S'_\mathrm{s,sp}(\x)$ in hand, the PSF model for each image is parametrized solely by $\beta_i$, which scales the speckle map.

\subsection{Subtraction Procedure}\label{subsec:sub_procedure}
We subtract the model PSF $\hat{S}_i$ from each image.  The map scale parameter $\beta_i$ and a registration offset are tuned to minimize subtraction residuals, which are rotated to a common frame and combined to form the image of the disk.

Special software masks are used in various stages of the subtraction process.  Diffraction spikes and the edge-on disk were masked during calculation of the radial profile.  The disk was also masked from individual frames before they were combined to form the speckle map.  Diffraction spikes, the focal plane spot, and the disk were excluded from consideration during the least-squares optimization of the PSF model subtraction.  Finally, diffraction spikes were masked from the subtraction residuals before they were combined to form the final disk image.

In summary, the sequence is to
\begin{enumerate}
\item register each frame to a fiducial position using stellar centroid estimates,
\item mask the diffraction spikes and the disk, then extract the radial profile about the fiducial point ($\rho_i$),
\item remove the profile and combine the residuals from each image to obtain the speckle map ($S'_\mathrm{s,sp}$),
\item construct a PSF model for each image, using a scaled speckle map ($\beta_i S'_\mathrm{s,sp}$) and the radial profile ($\rho_i$),
\item use an optimization method to subtract PSF model from image, solving for stellar position and speckle map scale ($\beta_i$),
\item repeat the process from step (1) using updated centroid estimates if not in final iteration,
\item mask diffraction spikes, rotate the subtraction residuals into the sky's frame, and combine to form the disk image estimate.
\end{enumerate}
We find that, for these data, three iterations of this process is enough for convergence to visually detectable levels.

\subsection{Technique Comparison}\label{subsec:sub_comparison}

The subtraction procedure described in the previous sections is compared with similar implementations in Figure~\ref{fig:subtraction_comparison}.  In panel \textit{(a)}, we follow the steps we outlined above.  In panel \textit{(b)}, we restrict the model fitting process so that the scale of the speckle map ($\beta_i$) is fixed and only offsets were optimized.  Finally, in \textit{(c)} we discard the PSF models altogether and fit the offset and scale of an average PSF.  No spike or object masking were used when combining residual images in \textit{(c)}.  In panel \textit{(d)} we display curves of annular \textit{rms} (corresponding to the photometric errors shown in Figure~\ref{fig:sbprof}), excluding the regions near the disk.  These curves show that fitting the speckle scale gives modest contrast gain ($\lesssim 0.1$\,mag), while larger gains are obtained when removing the radial profile of the observations prior to combination ($\sim 0.5$\,mag).

\section{MIDPLANE DETECTION SENSITIVITY}\label{ap:pt_src}

When quoting sensitivity limits for the detection of point sources around a star, it is common practice to measure the standard deviation of pixel values in concentric annuli, $\hat{\sigma}(r)$, and to set a scale to this curve with artificially inserted sources.   In this case, $\sigma$ represents a by-eye threshold.  Here we wish to extend this methodology for the case of sources within the disk.  To this end, we adopt a model for the sensitivity and solve for this model's parameters through visual detection of randomly inserted sources (both inside and outside of the disk) and maximum likelihood techniques.

We assume that the process of detecting a source can be characterized by zero-mean, normally distributed background noise fluctuations.  In this case, the background can be characterized by $\sigma$, and the probability of detecting a source (denoting $D=1$ for detection) of flux $f$ is given by
\begin{equation}
P(D=1|f,\sigma) = \frac{1}{2}\left[1+\mathrm{erf}\left(\frac{f}{2\sigma}\right)\right],
\end{equation}
which is the cumulative normal distribution function.  Our task is to determine $\sigma$, which we can assume is proportional to $\hat{\sigma}(r)$ in regions devoid of disk light.  Similarly, in the disk midplane, we expect an additional contribution to the background fluctuations from the disk light.  For our ground-based observations, these fluctuations are dominated by speckle noise (including Strehl fluctuations) rather than photon statistics, and therefore the variance of this contribution is proportional to the square of the disk brightness, $I^2$~\citep[e.g.][]{aime&soummer04,fitzgerald&graham06}.  We model the sensitivity in the midplane by
\begin{equation}
\sigma^2(\x, \mu, \xi) = \mu^2 \hat{\sigma}^2(\x) + \xi^2 I^2(\x).
\end{equation}
We fix $\xi=0$ to model the sensitivity in regions devoid of disk light.  We measure $\hat{\sigma}(r)$ with a sigma-clipped sample standard deviation of pixels in concentric annuli, excluding regions of disk emission.

We randomly generated positions and fluxes for artificial sources to be inserted in each of our ground-based images.  Two populations of sources were generated --- one set randomly distributed along the disk midplane, and another in the off-disk region.  The number of sources in each population was drawn from a Poisson distribution, and no blending of sources was allowed.  Fluxes were drawn from a log-uniform distribution about $\sigma(\x,\mu=1,\xi=1)$ in the midplane and $\sigma(\x,\mu=1,\xi=0)$ elsewhere.
After the computer inserted the randomly generated sources, the images were inspected and detections recorded.  False detections were ignored in our analysis.

\begin{figure}
\plotone{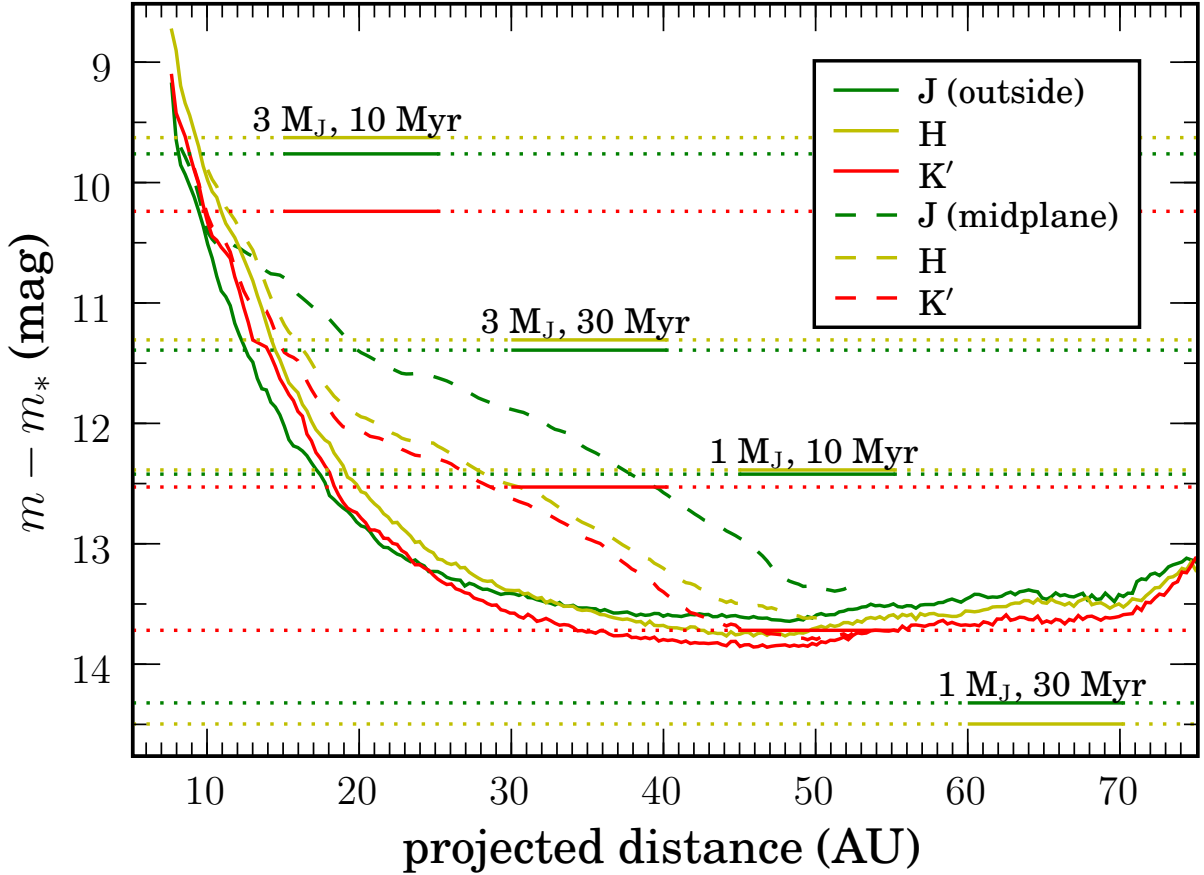}
\caption{Point source detection limits (5-$\sigma$).  The solid curves are for sources outside of the disk, while the dashed curves give limits for sources residing in the disk midplane.  In some places the midplane is more than a magnitude less sensitive.  No point sources were detected along the disk.  Horizontal lines indicate the predicted brightnesses for the model giant planets of~\citet{burrows_etal97}, at ages of 10 and 30\,Myr~\citep[\AUMic\ is $12^{+8}_{-4}$\,Myr;][]{barradoynavascues99}.  Initial conditions play a large role in the luminosity evolution of young $\sim1$\,$M_\mathrm{J}$ planets, and these ``hot start'' models represent brightness upper limits~\citep[e.g.][]{fortney_etal05, marley_etal07}.}\label{fig:sensitivity}
\end{figure}

\begin{figure}
\plottwo{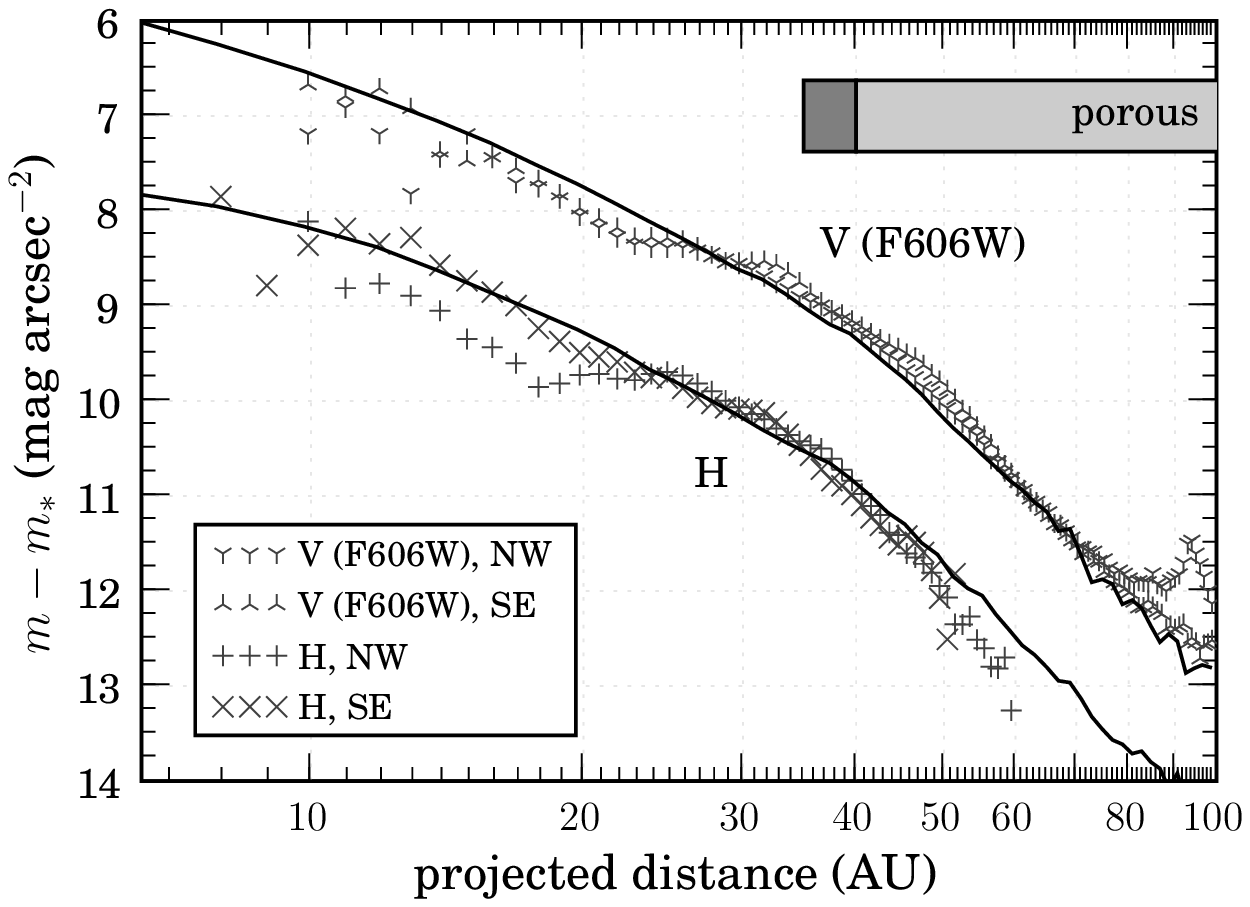}{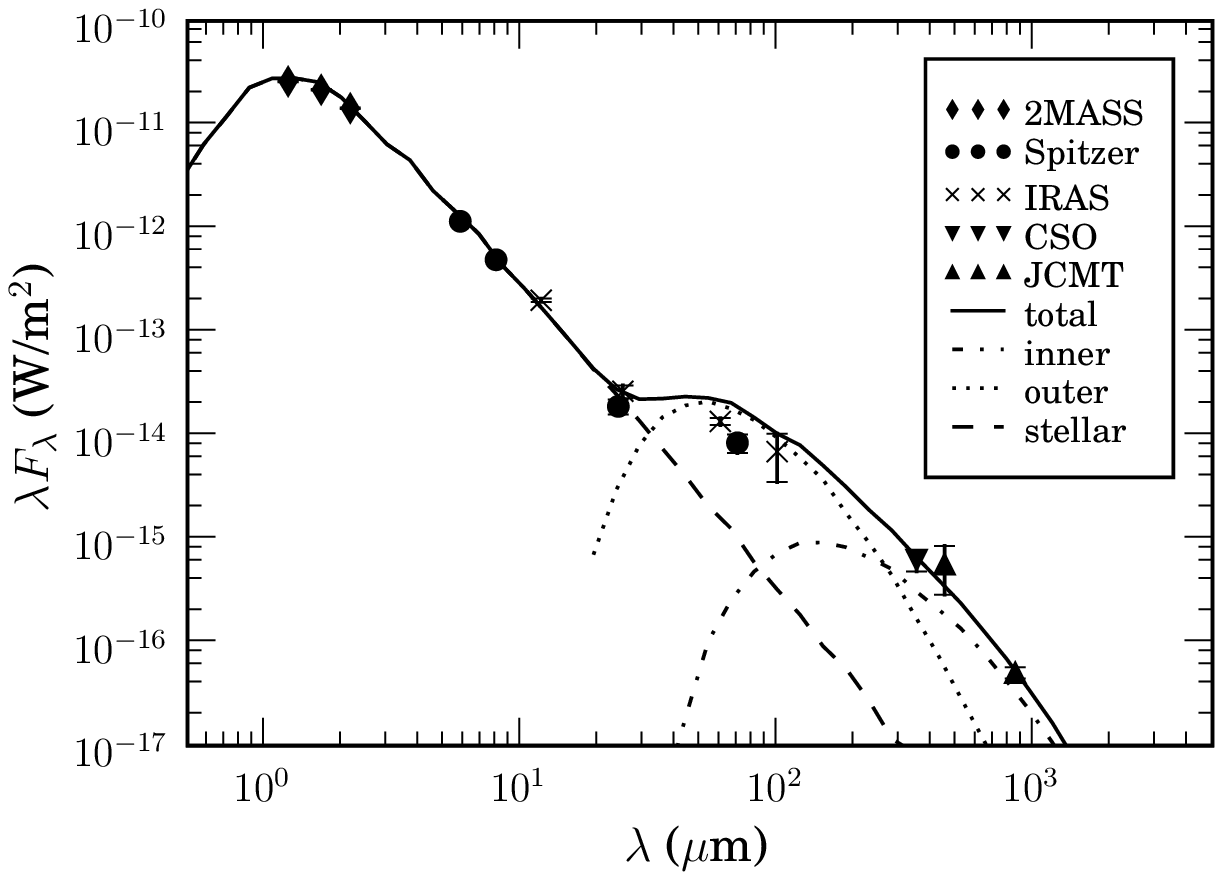}
\caption{A model fit to the scattered light and SED of the system.  \textit{(left)} The surface brightness profiles from Fig.~\ref{fig:sbprof} along with surface brightness profiles from the best-fit model.  The gray boxes above the profiles represent the grain locations in our model (cf. Table~\ref{tab:modelparms}); the dark region indicates the inner region of larger grains, while the smaller scatterers are in the lighter zone outside.  \textit{(right)} The model SED along with measured photometry of \AUMic.  In this model, the smaller grains are responsible for the bulk of the scattered light and the mid-IR emission, while the larger grains reproduce the long-wavelength end of the SED.}\label{fig:model}
\end{figure}

\begin{figure}
\plottwo{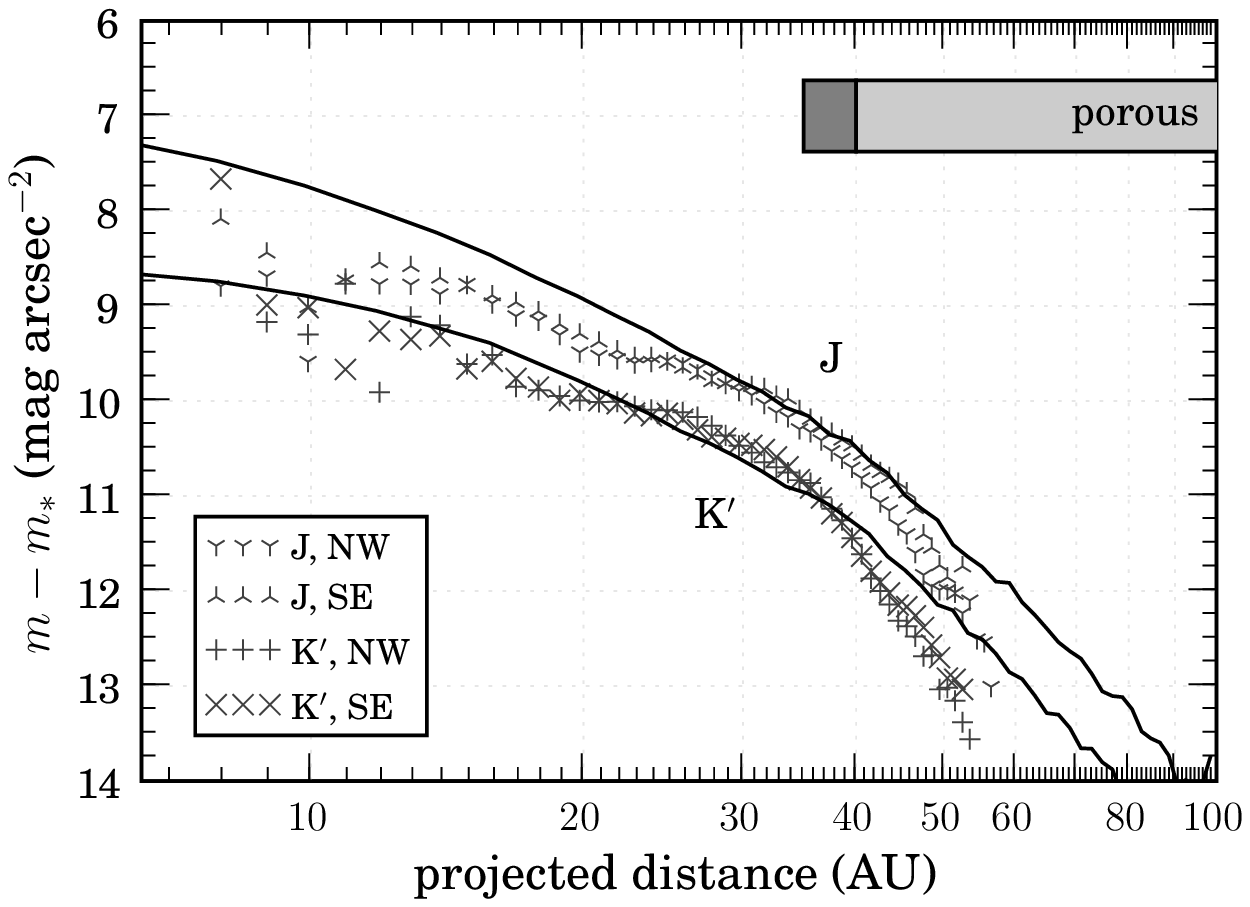}{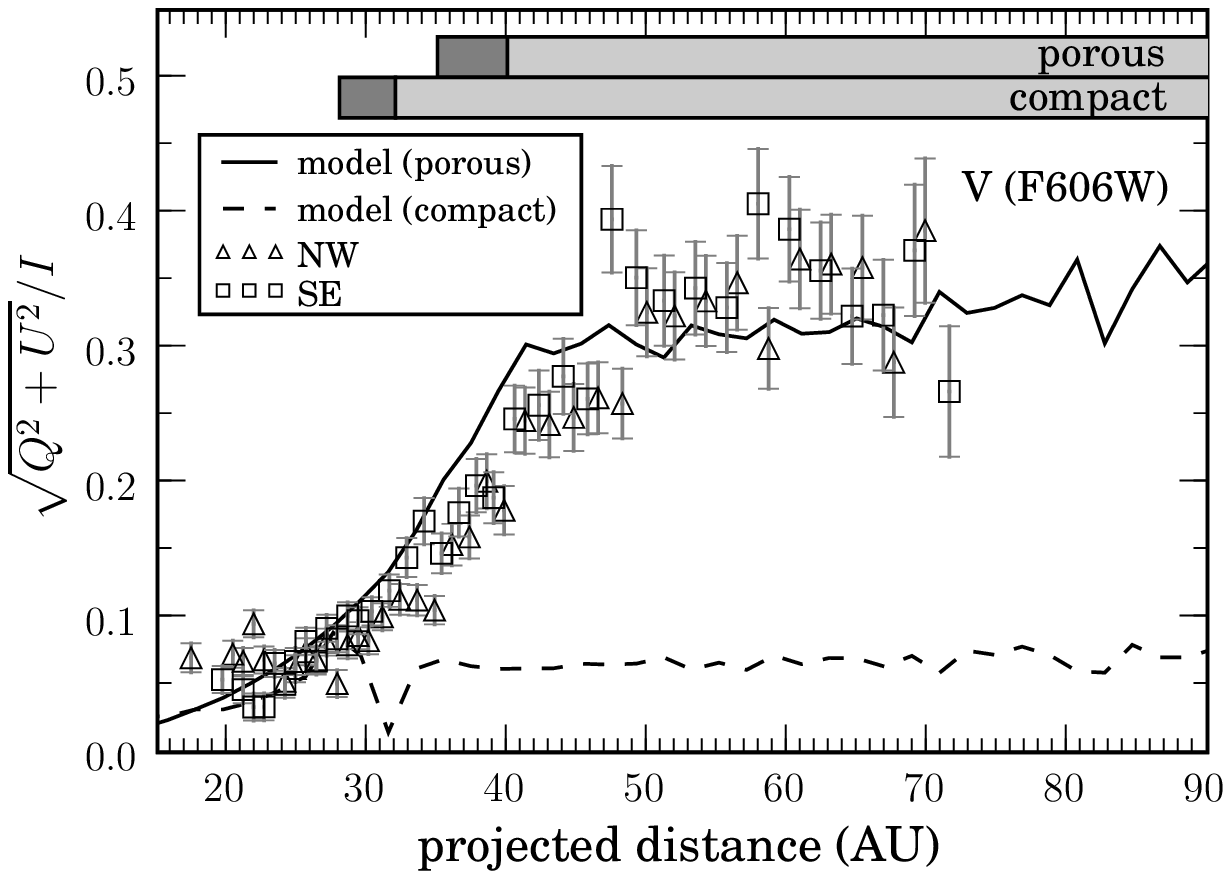}
\caption{\textit{(left)} A comparison of the surface brightness profiles from Fig.~\ref{fig:sbprof} to surface brightness profiles from the best-fit model in \bJ\ and \bKp\ bands.  These data were not used in the fitting process.  The differences in overall flux may be due to photometric calibration uncertainties~(\S\ref{subsec:calibration}), though the model clearly overestimates the emission in the outer zone in these bands.  More complicated models may resolve these discrepancies; a changing minimum grain size with radius can produce a color gradient.  \textit{(right)} The fraction of linear \bFW\ polarization produced by our model compared to measurements of~\citet{graham_etal07}, which include both systematic and random errors.  Compact, \micron-sized grains are ruled out by these data.}\label{fig:model_JK_pol}
\end{figure}

\begin{figure}
\plotone{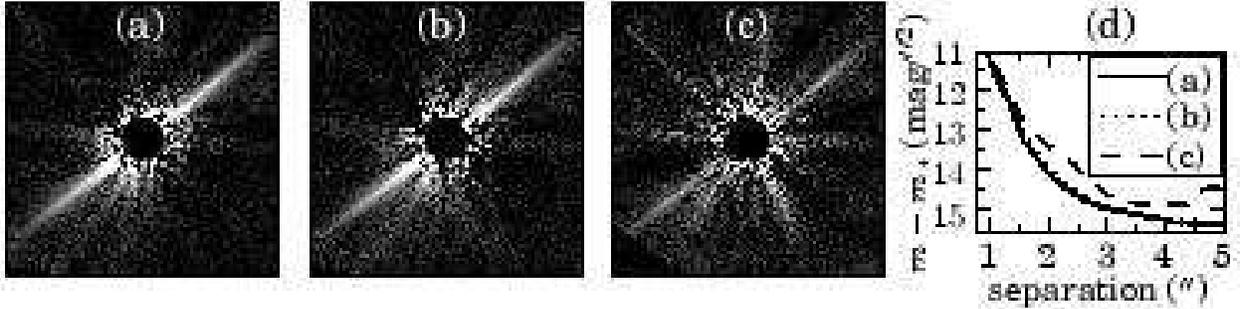}
\caption{A comparison of different methods of roll subtraction.  Panel \textit{(a)} shows the resulting image from the technique described in~\S\ref{subsec:psf_sub}.  Panel \textit{(b)} is the same as \textit{(a)}, except the scale of the speckle map is not optimized~(\S\ref{subsec:sub_procedure}).  For comparison, panel \textit{(c)} registers an average PSF to each image, rather than fitting a PSF model.  Each of these images is 8\arcsec\ on a side.  In \textit{(d)}, we show the annular \textit{rms} (excluding the disk) of photometry in 0\farcs 1$\,\times\,$0\farcs 5 apertures as a function of radius for subtractions shown in \textit{(a)-(c)}.  At a radius of 2\arcsec, curve \textit{(b)} is 0.56\,mag more sensitive than \textit{(c)}, while \textit{(a)} is 0.16\,mag more sensitive than \textit{(b)}.  The gain in contrast when subtracting the profile highlights the suitability of the technique for edge-on disks.}\label{fig:subtraction_comparison}
\end{figure}

With these data in hand, the problem reduces to finding the most likely model parameters ($\mu$, $\xi$) given the set of artificial source positions, fluxes, and detections $\left\{\x_i, f_i, D_i\right\}$.  Our sensitivity curves will use the model parameters which maximize the probability density $p(\mu, \xi | \left\{\x_i, f_i, D_i\right\})$.  This quantity can be re-written with Bayes' Theorem,
\begin{eqnarray}
p(\mu, \xi | \left\{\x_i, f_i, D_i\right\}) &\propto& p(\mu, \xi) p(\left\{\x_i, f_i, D_i\right\} | \mu, \xi), \\
&=& p(\mu, \xi) \prod_i P(D_i | \x_i, f_i, \mu, \xi) p(\x_i, f_i), \\
&\propto& p(\mu, \xi) \prod_i P(D_i | \x_i, f_i, \mu, \xi).
\end{eqnarray}
The final step is valid since we are free to choose the position and flux distribution of independent artificial sources without regard to the model parameters.  Our prior information on the model parameters is represented by $p(\mu, \xi)$, and we assume uniform distributions which are also independent of $\mu, \xi$.  We maximized the logarithm of this function with respect to $\mu$ and $\xi$ for each of our \bJ\bH\bKp\ images, to obtain best-fit 1-$\sigma$ sensitivities, and we show our 5-$\sigma$ point source detection sensitivity as a function of separation from the star in Figure~\ref{fig:sensitivity}.  Midplane sensitivities are averaged over both ansae.  We show predicted planet brightnesses using models of~\citet{burrows_etal97} at different ages, allowing inference of detection limits for planet mass.

\bibliography{}

\begin{thebibliography}{79}

\bibitem[{{Aime} \& {Soummer}(2004)}]{aime&soummer04}
{Aime}, C. \& {Soummer}, R. 2004, \apjl, 612, L85

\bibitem[{{Artymowicz}(1988)}]{artymowicz88}
{Artymowicz}, P. 1988, \apjl, 335, L79

\bibitem[{{Augereau} \& {Beust}(2006)}]{augereau&beust06}
{Augereau}, J.~C. \& {Beust}, H. 2006, \aap, 455, 987

\bibitem[{{Augereau} {et~al.}(2001){Augereau}, {Nelson}, {Lagrange},  {Papaloizou}, \& {Mouillet}}]{augereau_etal01}
{Augereau}, J.~C., {Nelson}, R.~P., {Lagrange}, A.~M., {Papaloizou}, J.~C.~B.,  \& {Mouillet}, D. 2001, \aap, 370, 447

\bibitem[{{Backman} \& {Paresce}(1993)}]{backman&paresce93}
{Backman}, D.~E. \& {Paresce}, F. 1993, in Protostars and Planets III,  1253--1304

\bibitem[{{Barrado y Navascu{\'e}s} {et~al.}(1999){Barrado y Navascu{\'e}s},  {Stauffer}, {Song}, \& {Caillault}}]{barradoynavascues99}
{Barrado y Navascu{\'e}s}, D., {Stauffer}, J.~R., {Song}, I., \& {Caillault},  J.-P. 1999, \apjl, 520, L123

\bibitem[{{Bryden} {et~al.}(2006){Bryden}, {Beichman}, {Trilling}, {Rieke},  {Holmes}, {Lawler}, {Stapelfeldt}, {Werner}, {Gautier}, {Blaylock}, {Gordon},  {Stansberry}, \& {Su}}]{bryden_etal06}
{Bryden}, G., {et~al.} 2006, \apj, 636, 1098

\bibitem[{{Burrows} {et~al.}(1997){Burrows}, {Marley}, {Hubbard}, {Lunine},  {Guillot}, {Saumon}, {Freedman}, {Sudarsky}, \& {Sharp}}]{burrows_etal97}
{Burrows}, A., {et~al.} 1997, \apj, 491,  856

\bibitem[{{Chen} {et~al.}(2005){Chen}, {Patten}, {Werner}, {Dowell},  {Stapelfeldt}, {Song}, {Stauffer}, {Blaylock}, {Gordon}, \&  {Krause}}]{chen_etal05}
{Chen}, C.~H., {et~al.} 2005, \apj, 634, 1372

\bibitem[{{Chen} {et~al.}(2006){Chen}, {Sargent}, {Bohac}, {Kim},  {Leibensperger}, {Jura}, {Najita}, {Forrest}, {Watson}, {Sloan}, \&  {Keller}}]{chen_etal06}
{Chen}, C.~H., {et~al.} 2006, \apjs, 166, 351

\bibitem[{{Cutispoto} {et~al.}(2003){Cutispoto}, {Messina}, \&  {Rodon{\`o}}}]{cutispoto_etal03}
{Cutispoto}, G., {Messina}, S., \& {Rodon{\`o}}, M. 2003, \aap, 400, 659

\bibitem[{{Decin} {et~al.}(2003){Decin}, {Dominik}, {Waters}, \&  {Waelkens}}]{decin_etal03}
{Decin}, G., {Dominik}, C., {Waters}, L.~B.~F.~M., \& {Waelkens}, C. 2003,  \apj, 598, 636

\bibitem[{{Dohnanyi}(1969)}]{dohnanyi69}
{Dohnanyi}, J.~S. 1969, \jgr, 74, 2431

\bibitem[{{Draine} \& {Li}(2001)}]{draine&li01}
{Draine}, B.~T. \& {Li}, A. 2001, \apj, 551, 807

\bibitem[{{Elias} {et~al.}(1982){Elias}, {Frogel}, {Matthews}, \&  {Neugebauer}}]{elias_etal82}
{Elias}, J.~H., {Frogel}, J.~A., {Matthews}, K., \& {Neugebauer}, G. 1982, \aj,  87, 1029

\bibitem[{{Fitzgerald} \& {Graham}(2006)}]{fitzgerald&graham06}
{Fitzgerald}, M.~P. \& {Graham}, J.~R. 2006, \apj, 637, 541

\bibitem[{{Fortney} {et~al.}(2005){Fortney}, {Marley}, {Hubickyj},  {Bodenheimer}, \& {Lissauer}}]{fortney_etal05}
{Fortney}, J.~J., {Marley}, M.~S., {Hubickyj}, O., {Bodenheimer}, P., \&  {Lissauer}, J.~J. 2005, Astronomische Nachrichten, 326, 925

\bibitem[{{Gliese} \& {Jahreiss}(1995)}]{gliese&jahreiss95}
{Gliese}, W. \& {Jahreiss}, H. 1995, VizieR Online Data Catalog, 5070, 0

\bibitem[{{Golimowski} {et~al.}(2006){Golimowski}, {Ardila}, {Krist},  {Clampin}, {Ford}, {Illingworth}, {Bartko}, {Ben{\'{\i}}tez}, {Blakeslee},  {Bouwens}, {Bradley}, {Broadhurst}, {Brown}, {Burrows}, {Cheng}, {Cross},  {Demarco}, {Feldman}, {Franx}, {Goto}, {Gronwall}, {Hartig}, {Holden},  {Homeier}, {Infante}, {Jee}, {Kimble}, {Lesser}, {Martel}, {Mei},  {Menanteau}, {Meurer}, {Miley}, {Motta}, {Postman}, {Rosati}, {Sirianni},  {Sparks}, {Tran}, {Tsvetanov}, {White}, {Zheng}, \&  {Zirm}}]{golimowski_etal06}
{Golimowski}, D.~A., {et~al.} 2006, \aj, 131, 3109

\bibitem[{{Graham} {et~al.}(2007){Graham}, {Kalas}, \&  {Matthews}}]{graham_etal07}
{Graham}, J.~R., {Kalas}, P.~G., \& {Matthews}, B.~C. 2007, \apj, 654, 595

\bibitem[{{Grigorieva} {et~al.}(2007){Grigorieva}, {Artymowicz}, \&  {Th{\'e}bault}}]{grigorieva_etal07}
{Grigorieva}, A., {Artymowicz}, P., \& {Th{\'e}bault}, P. 2007, \aap, 461, 537

\bibitem[{{Haisch} {et~al.}(2001){Haisch}, {Lada}, \& {Lada}}]{haisch_etal01}
{Haisch}, Jr., K.~E., {Lada}, E.~A., \& {Lada}, C.~J. 2001, \apjl, 553, L153

\bibitem[{{Hauschildt} {et~al.}(1999){Hauschildt}, {Allard}, \&  {Baron}}]{hauschildt_etal99}
{Hauschildt}, P.~H., {Allard}, F., \& {Baron}, E. 1999, \apj, 512, 377

\bibitem[{{Joy} \& {Abt}(1974)}]{joy&abt74}
{Joy}, A.~H. \& {Abt}, H.~A. 1974, \apjs, 28, 1

\bibitem[{{Kalas}(2005)}]{kalas05}
{Kalas}, P. 2005, \apjl, 635, L169

\bibitem[{{Kalas} {et~al.}(2007){Kalas}, {Fitzgerald}, \&  {Graham}}]{kalas_etal07}
{Kalas}, P., {Fitzgerald}, M.~P., \& {Graham}, J.~R. 2007, \apjl, 661, L85

\bibitem[{{Kalas} {et~al.}(2006){Kalas}, {Graham}, {Clampin}, \&  {Fitzgerald}}]{kalas_etal06}
{Kalas}, P., {Graham}, J.~R., {Clampin}, M.~C., \& {Fitzgerald}, M.~P. 2006,  \apjl, 637, L57

\bibitem[{{Kalas} {et~al.}(2004){Kalas}, {Liu}, \& {Matthews}}]{kalas_etal04}
{Kalas}, P., {Liu}, M.~C., \& {Matthews}, B.~C. 2004, Science, 303, 1990

\bibitem[{{Keenan}(1983)}]{keenan83}
{Keenan}, P.~C. 1983, Bulletin d'Information du Centre de Donnees Stellaires,  24, 19

\bibitem[{{Kenyon} \& {Bromley}(2004)}]{kenyon&bromley04}
{Kenyon}, S.~J. \& {Bromley}, B.~C. 2004, \aj, 127, 513

\bibitem[{{Krisciunas} {et~al.}(1987){Krisciunas}, {Sinton}, {Tholen},  {Tokunaga}, {Golisch}, {Griep}, {Kaminski}, {Impey}, \&  {Christian}}]{krisciunas_etal87}
{Krisciunas}, K., {et~al.} 1987, \pasp, 99,  887

\bibitem[{{Krist} {et~al.}(2005){Krist}, {Ardila}, {Golimowski}, {Clampin},  {Ford}, {Illingworth}, {Hartig}, {Bartko}, {Ben{\'{\i}}tez}, {Blakeslee},  {Bouwens}, {Bradley}, {Broadhurst}, {Brown}, {Burrows}, {Cheng}, {Cross},  {Demarco}, {Feldman}, {Franx}, {Goto}, {Gronwall}, {Holden}, {Homeier},  {Infante}, {Kimble}, {Lesser}, {Martel}, {Mei}, {Menanteau}, {Meurer},  {Miley}, {Motta}, {Postman}, {Rosati}, {Sirianni}, {Sparks}, {Tran},  {Tsvetanov}, {White}, \& {Zheng}}]{krist_etal05}
{Krist}, J.~E., {et~al.} 2005, \aj,  129, 1008

\bibitem[{{Krist} \& {Hook}(2004)}]{krist&hook04}
{Krist}, J.~E. \& {Hook}, R. 2004, {The Tiny Tim User's Guide, Version 6.3}  (Baltimore: STScI)

\bibitem[{{Kuchner} \& {Holman}(2003)}]{kuchner&holman03}
{Kuchner}, M.~J. \& {Holman}, M.~J. 2003, \apj, 588, 1110

\bibitem[{{Lagrange} {et~al.}(2000){Lagrange}, {Backman}, \&  {Artymowicz}}]{lagrange_etal00}
{Lagrange}, A.-M., {Backman}, D.~E., \& {Artymowicz}, P. 2000, Protostars and  Planets IV, 639

\bibitem[{{Lecavelier Des Etangs} {et~al.}(1996){Lecavelier Des Etangs},  {Vidal-Madjar}, \& {Ferlet}}]{lecavelier_etal96}
{Lecavelier Des Etangs}, A., {Vidal-Madjar}, A., \& {Ferlet}, R. 1996, \aap,  307, 542

\bibitem[{{Li} \& {Greenberg}(1998)}]{li&greenberg98}
{Li}, A. \& {Greenberg}, J.~M. 1998, \aap, 331, 291

\bibitem[{{Linsky} {et~al.}(1982){Linsky}, {Bornmann}, {Carpenter}, {Hege},  {Wing}, {Giampapa}, \& {Worden}}]{linsky_etal82}
{Linsky}, J.~L., {Bornmann}, P.~L., {Carpenter}, K.~G., {Hege}, E.~K., {Wing},  R.~F., {Giampapa}, M.~S., \& {Worden}, S.~P. 1982, \apj, 260, 670

\bibitem[{{Liou} \& {Zook}(1999)}]{liou&zook99}
{Liou}, J.-C. \& {Zook}, H.~A. 1999, \aj, 118, 580

\bibitem[{{Liu}(2004)}]{liu04}
{Liu}, M.~C. 2004, Science, 305, 1442

\bibitem[{{Liu} {et~al.}(2004){Liu}, {Matthews}, {Williams}, \&  {Kalas}}]{liu_etal04}
{Liu}, M.~C., {Matthews}, B.~C., {Williams}, J.~P., \& {Kalas}, P.~G. 2004,  \apj, 608, 526

\bibitem[{{Lloyd} \& {Sivaramakrishnan}(2005)}]{lloyd&sivaramakrishnan05}
{Lloyd}, J.~P. \& {Sivaramakrishnan}, A. 2005, \apj, 621, 1153

\bibitem[{{Marley} {et~al.}(2007){Marley}, {Fortney}, {Hubickyj},  {Bodenheimer}, \& {Lissauer}}]{marley_etal07}
{Marley}, M.~S., {Fortney}, J.~J., {Hubickyj}, O., {Bodenheimer}, P., \&  {Lissauer}, J.~J. 2007, \apj, 655, 541

\bibitem[{{Marois} {et~al.}(2006){Marois}, {Lafreni{\`e}re},  {Doyon}, {Macintosh}, \& {Nadeau}}]{marois_etal06}
{Marois}, C., {Lafreni{\`e}re}, D., {Doyon}, R., {Macintosh}, B.,  \& {Nadeau}, D. 2006, \apj, 641, 556

\bibitem[{{Masciadri} {et~al.}(2005){Masciadri}, {Mundt}, {Henning}, {Alvarez},  \& {Barrado y Navascu{\'e}s}}]{masciadri_etal05}
{Masciadri}, E., {Mundt}, R., {Henning}, T., {Alvarez}, C., \& {Barrado y  Navascu{\'e}s}, D. 2005, \apj, 625, 1004

\bibitem[{{Mathioudakis} \& {Doyle}(1991)}]{mathioudakis&doyle91}
{Mathioudakis}, M. \& {Doyle}, J.~G. 1991, \aap, 244, 433

\bibitem[{{Mathis} \& {Whiffen}(1989)}]{mathis&whiffen89}
{Mathis}, J.~S. \& {Whiffen}, G. 1989, \apj, 341, 808

\bibitem[{{Metchev} {et~al.}(2005){Metchev}, {Eisner}, {Hillenbrand}, \&  {Wolf}}]{metchev_etal05}
{Metchev}, S.~A., {Eisner}, J.~A., {Hillenbrand}, L.~A., \& {Wolf}, S. 2005,  \apj, 622, 451

\bibitem[{{Meyer} {et~al.}(2007){Meyer}, {Backman}, {Weinberger}, \&  {Wyatt}}]{meyer_etal07}
{Meyer}, M.~R., {Backman}, D.~E., {Weinberger}, A.~J., \& {Wyatt}, M.~C. 2007,  in Protostars and Planets V, ed. B.~{Reipurth}, D.~{Jewitt}, \& K.~{Keil},  573--588

\bibitem[{{Moro-Mart{\'{\i}}n} \& {Malhotra}(2005)}]{moro-martin&malhotra05}
{Moro-Mart{\'{\i}}n}, A. \& {Malhotra}, R. 2005, \apj, 633, 1150

\bibitem[{{Moro-Mart{\'{\i}}n} {et~al.}(2005){Moro-Mart{\'{\i}}n}, {Wolf}, \&  {Malhotra}}]{moro-martin_etal05}
{Moro-Mart{\'{\i}}n}, A., {Wolf}, S., \& {Malhotra}, R. 2005, \apj, 621, 1079

\bibitem[{{Neuh{\"a}user} {et~al.}(2003){Neuh{\"a}user}, {Guenther}, {Alves},  {Hu{\'e}lamo}, {Ott}, \& {Eckart}}]{neuhauser_etal03}
{Neuh{\"a}user}, R., {Guenther}, E.~W., {Alves}, J., {Hu{\'e}lamo}, N., {Ott},  T., \& {Eckart}, A. 2003, Astronomische Nachrichten, 324, 535

\bibitem[{{Okamoto} {et~al.}(2004){Okamoto}, {Kataza}, {Honda}, {Yamashita},  {Onaka}, {Watanabe}, {Miyata}, {Sako}, {Fujiyoshi}, \&  {Sakon}}]{okamoto_etal04}
{Okamoto}, Y.~K., {et~al.} 2004, \nat, 431, 660

\bibitem[{{Ozernoy} {et~al.}(2000){Ozernoy}, {Gorkavyi}, {Mather}, \&  {Taidakova}}]{ozernoy_etal00}
{Ozernoy}, L.~M., {Gorkavyi}, N.~N., {Mather}, J.~C., \& {Taidakova}, T.~A.  2000, \apjl, 537, L147

\bibitem[{{Pantin} {et~al.}(1997){Pantin}, {Lagage}, \&  {Artymowicz}}]{pantin_etal97}
{Pantin}, E., {Lagage}, P.~O., \& {Artymowicz}, P. 1997, \aap, 327, 1123

\bibitem[{{Perryman} {et~al.}(1997){Perryman}, {Lindegren}, {Kovalevsky},  {Hoeg}, {Bastian}, {Bernacca}, {Cr{\'e}z{\'e}}, {Donati}, {Grenon}, {van  Leeuwen}, {van der Marel}, {Mignard}, {Murray}, {Le Poole}, {Schrijver},  {Turon}, {Arenou}, {Froeschl{\'e}}, \& {Petersen}}]{perryman_etal97}
{Perryman}, M.~A.~C., {et~al.} 1997, \aap, 323, L49

\bibitem[{{Persson} {et~al.}(1998){Persson}, {Murphy}, {Krzeminski}, {Roth}, \&  {Rieke}}]{persson_etal98}
{Persson}, S.~E., {Murphy}, D.~C., {Krzeminski}, W., {Roth}, M., \& {Rieke},  M.~J. 1998, \aj, 116, 2475

\bibitem[{{Pinte} {et~al.}(2006){Pinte}, {M{\'e}nard}, {Duch{\^e}ne}, \&  {Bastien}}]{pinte_etal06}
{Pinte}, C., {M{\'e}nard}, F., {Duch{\^e}ne}, G., \& {Bastien}, P. 2006, \aap,  459, 797

\bibitem[{{Plavchan} {et~al.}(2005){Plavchan}, {Jura}, \&  {Lipscy}}]{plavchan_etal05}
{Plavchan}, P., {Jura}, M., \& {Lipscy}, S.~J. 2005, \apj, 631, 1161

\bibitem[{{Roberge} {et~al.}(2005){Roberge}, {Weinberger}, {Redfield}, \&  {Feldman}}]{roberge_etal05}
{Roberge}, A., {Weinberger}, A.~J., {Redfield}, S., \& {Feldman}, P.~D. 2005,  \apjl, 626, L105

\bibitem[{{Robinson} {et~al.}(2001){Robinson}, {Linsky}, {Woodgate}, \&  {Timothy}}]{robinson_etal01}
{Robinson}, R.~D., {Linsky}, J.~L., {Woodgate}, B.~E., \& {Timothy}, J.~G.  2001, \apj, 554, 368

\bibitem[{{Roques} {et~al.}(1994){Roques}, {Scholl}, {Sicardy}, \&  {Smith}}]{roques_etal94}
{Roques}, F., {Scholl}, H., {Sicardy}, B., \& {Smith}, B.~A. 1994, Icarus, 108,  37

\bibitem[{{Sheehy} {et~al.}(2006){Sheehy}, {McCrady}, \&  {Graham}}]{sheehy_etal06}
{Sheehy}, C.~D., {McCrady}, N., \& {Graham}, J.~R. 2006, \apj, 647, 1517

\bibitem[{{Skrutskie} {et~al.}(2006){Skrutskie}, {Cutri}, {Stiening},  {Weinberg}, {Schneider}, {Carpenter}, {Beichman}, {Capps}, {Chester},  {Elias}, {Huchra}, {Liebert}, {Lonsdale}, {Monet}, {Price}, {Seitzer},  {Jarrett}, {Kirkpatrick}, {Gizis}, {Howard}, {Evans}, {Fowler}, {Fullmer},  {Hurt}, {Light}, {Kopan}, {Marsh}, {McCallon}, {Tam}, {Van Dyk}, \&  {Wheelock}}]{skrutskie_etal06}
{Skrutskie}, M.~F., {et~al.} 2006, \aj, 131, 1163

\bibitem[{{Song} {et~al.}(2002){Song}, {Weinberger}, {Becklin}, {Zuckerman}, \&  {Chen}}]{song_etal02}
{Song}, I., {Weinberger}, A.~J., {Becklin}, E.~E., {Zuckerman}, B., \& {Chen},  C. 2002, \aj, 124, 514

\bibitem[{{Strubbe} \& {Chiang}(2006)}]{strubbe&chiang06}
{Strubbe}, L.~E. \& {Chiang}, E.~I. 2006, \apj, 648, 652

\bibitem[{{Telesco} {et~al.}(2005){Telesco}, {Fisher}, {Wyatt}, {Dermott},  {Kehoe}, {Novotny}, {Mari{\~n}as}, {Radomski}, {Packham}, {De Buizer}, \&  {Hayward}}]{telesco_etal05}
{Telesco}, C.~M., {et~al.} 2005, \nat, 433, 133

\bibitem[{{Thommes} \& {Lissauer}(2003)}]{thommes&lissauer03}
{Thommes}, E.~W. \& {Lissauer}, J.~J. 2003, \apj, 597, 566

\bibitem[{{Torres} \& {Ferraz Mello}(1973)}]{torres&ferrazmello73}
{Torres}, C.~A.~O. \& {Ferraz Mello}, S. 1973, \aap, 27, 231

\bibitem[{{V{\' e}ran} {et~al.}(1997){V{\' e}ran}, {Rigaut}, {Ma{\^ i}tre}, \&  {Rouan}}]{veran_etal97}
{V{\' e}ran}, J.-P., {Rigaut}, F., {Ma{\^ i}tre}, H., \& {Rouan}, D. 1997, 14,  3057

\bibitem[{{Voshchinnikov} {et~al.}(2006){Voshchinnikov}, {Il'in}, {Henning}, \&  {Dubkova}}]{voschinnikov_etal06}
{Voshchinnikov}, N.~V., {Il'in}, V.~B., {Henning}, T., \& {Dubkova}, D.~N.  2006, \aap, 445, 167

\bibitem[{{Voshchinnikov} \& {Mathis}(1999)}]{voshchinnikov&mathis99}
{Voshchinnikov}, N.~V. \& {Mathis}, J.~S. 1999, \apj, 526, 257

\bibitem[{{Wahhaj} {et~al.}(2003){Wahhaj}, {Koerner}, {Ressler}, {Werner},  {Backman}, \& {Sargent}}]{wahhaj_etal03}
{Wahhaj}, Z., {Koerner}, D.~W., {Ressler}, M.~E., {Werner}, M.~W., {Backman},  D.~E., \& {Sargent}, A.~I. 2003, \apjl, 584, L27

\bibitem[{{Weinberger} {et~al.}(2003){Weinberger}, {Becklin}, \&  {Zuckerman}}]{weinberger_etal03}
{Weinberger}, A.~J., {Becklin}, E.~E., \& {Zuckerman}, B. 2003, \apjl, 584, L33

\bibitem[{{Wyatt}(2006)}]{wyatt06}
{Wyatt}, M.~C. 2006, \apj, 639, 1153

\bibitem[{{Wyatt} {et~al.}(1999){Wyatt}, {Dermott}, {Telesco}, {Fisher},  {Grogan}, {Holmes}, \& {Pi{\~n}a}}]{wyatt_etal99}
{Wyatt}, M.~C., {Dermott}, S.~F., {Telesco}, C.~M., {Fisher}, R.~S., {Grogan},  K., {Holmes}, E.~K., \& {Pi{\~n}a}, R.~K. 1999, \apj, 527, 918

\bibitem[{{Zuckerman}(2001)}]{zuckerman01}
{Zuckerman}, B. 2001, \araa, 39, 549

\bibitem[{{Zuckerman} {et~al.}(1995){Zuckerman}, {Forveille}, \&  {Kastner}}]{zuckerman_etal95}
{Zuckerman}, B., {Forveille}, T., \& {Kastner}, J.~H. 1995, \nat, 373, 494

\bibitem[{{Zuckerman} {et~al.}(2001){Zuckerman}, {Song}, {Bessell}, \&  {Webb}}]{zuckerman_etal01}
{Zuckerman}, B., {Song}, I., {Bessell}, M.~S., \& {Webb}, R.~A. 2001, \apjl,  562, L87

\end{thebibliography}

\clearpage

\begin{deluxetable}{llcrc}
\tablewidth{0pc}
\tablecaption{Observations of \AUMic\label{tab:obs}}
\tablehead{\colhead{night} & \colhead{band} & \colhead{$r_\mathrm{mask}$ (\arcsec)} & \colhead{$T$ (s)} & \colhead{$\Delta$PA (\degr)}}
\startdata
2004 Aug 29 & \bJ  & 0.75 &  600 & 16.9 \\
            & \bH  & 0.75 &  690 & 49.8 \\
            & \bKp & 0.75 & 1110 & 40.5 \\
2004 Aug 30 & \bJ  & 0.50 &  570 & 39.0 \\
            & \bH  & 0.50 &  600 & 34.5 \\
            & \bKp & 0.50 & 1260 & 45.6 \\
            & \bLp & 0.50 &  720 & 43.6 \\
\enddata
\tablecomments{The radius of the coronagraphic focal plane mask is given by $r_\mathrm{mask}$.  Filters were cycled after a few short exposures in each band.  Here, $T$ is the total integration time and $\Delta$PA is the total amount of field rotation over the course of the exposure sequence in that band.}
\end{deluxetable}

\begin{deluxetable}{ccr@{ $\pm$ }lr@{ $\pm$ }l}
\tablewidth{0pc}
\tablecaption{Midplane Surface Brightness Power-Law Indices\label{tab:power_laws}}
\tablecolumns{6}
\tablehead{\colhead{band}&\colhead{ansa}&\multicolumn{4}{c}{fit domain}\\
\colhead{}&\colhead{}&\multicolumn{2}{c}{15--32\,AU}&\multicolumn{2}{c}{32--60\tablenotemark{\dag}\,AU}}
\startdata
\bFW & NW  &  1.46 & 0.09 &  3.00 & 0.13 \\
& SE  &  1.53 & 0.08 &  2.97 & 0.12 \\
& avg.  &  1.49 & 0.09 &  2.99 & 0.12 \\[5pt]
\bJ & NW  &  1.21 & 0.10 &  3.95 & 0.19 \\
& SE  &  1.34 & 0.12 &  3.27 & 0.19 \\
& avg.  &  1.27 & 0.11 &  3.61 & 0.19 \\[5pt]
\bH & NW  &  1.19 & 0.19 &  3.84 & 0.13 \\
& SE  &  1.58 & 0.15 &  3.47 & 0.17 \\
& avg.  &  1.39 & 0.17 &  3.66 & 0.15 \\[5pt]
\bKp & NW  &  1.24 & 0.13 &  5.16 & 0.23 \\
& SE  &  1.09 & 0.10 &  4.63 & 0.16 \\
& avg.  &  1.17 & 0.12 &  4.90 & 0.20 \\[5pt]
\enddata
 
\tablenotetext{\dag}{Maximum projected distance used in fit; the domain may be further restricted by the availability of data in Fig.~\ref{fig:sbprof}.}
\tablecomments{Power-law indices $\alpha$, calculated by converting the the midplane surface brightness profiles (Fig.~\ref{fig:sbprof}) to flux units and fitting $f(b)\propto b^{-\alpha}$.  Formal 1-$\sigma$ errors are scaled by $\sqrt{\chi^2_\nu}$ when this quantity is $>1$.  These errors are lower limits, as both systematic errors and measurement covariance have been ignored.  Entries marked ``avg.'' are computed by averaging $\alpha$ and its variance over both ansae.}
\end{deluxetable}

\begin{deluxetable}{ccccccc}
\tablewidth{0pc}
\tablecaption{Comparison of Disk Features\label{tab:struct}}
\tablehead{\colhead{Label}&\colhead{NW}&\colhead{SE}&\multicolumn{4}{c}{Location (AU)}\\
\colhead{}&\colhead{}&\colhead{}&\colhead{\citetalias{liu04}}&\colhead{\citetalias{krist_etal05}}&\colhead{\citetalias{metchev_etal05}}&\colhead{this work}}
\startdata
A & $\uparrow$   & $\uparrow$   & 25     & 26 & 22     & 25  \\
B & \nodata      & $\downarrow$ & 28.5   & 29 & 26     & 28  \\
C & \nodata      & $\uparrow$   & 31     & 33 & 32     & 32  \\
D & $\uparrow$   & $\downarrow$ & 37     & 37 & 38     & 37  \\
E & $\downarrow$ & $\downarrow$ & \ldots & 23 & \ldots & 21  \\
\enddata
\tablecomments{Features are marked in Figure~\ref{fig:substructure}.  Arrows denote localized enhancements ($\uparrow$) and deficits ($\downarrow$) of disk brightness.  Feature locations in~\citetalias{krist_etal05} are seen in \bFW, while~\citetalias{liu04} and~\citetalias{metchev_etal05} are \bH-band.  Average feature locations in the near-IR data from this \doctype\ are in the final column, based on visual positions in Fig.~\ref{fig:substructure} with uncertainties of approximately $\pm 0.5$\,AU.  We newly designate the close-in feature E here.}
\end{deluxetable}

\begin{deluxetable}{lr@{--}lr@{$\times$}lr@{--}lrr}
\tablewidth{0pc}
\tablecaption{Best-Fit Model Parameters\label{tab:modelparms}}
\tablehead{\colhead{region}&\multicolumn{2}{c}{range (AU)}&\multicolumn{2}{c}{dust mass ($M_\earth$)}&\multicolumn{2}{c}{$a$}&\colhead{$\gamma$}&\colhead{$p$}}
\startdata
\sidehead{porous~\citep{mathis&whiffen89}}
inner & 35&40 & 1.0&$10^{-2}$ & 3&6\,mm & -3.5\tablenotemark{\ddag} & +1.5 \\
outer & 40&300 & 2.3&$10^{-4}$ & 0.05&3.0\,\micron & -4.1 & -2.5 \\
\sidehead{compact~\citep{draine&li01}}
inner & 28&32 & 2.7&$10^{-2}$ & 1&2\,mm & -3.5\tablenotemark{\ddag} & +1.5 \\
outer & 32&300 & 2.7&$10^{-4}$ & 0.15&50.0\,\micron & -4.1 & -2.5 \\
\enddata
\tablenotetext{\ddag}{Fixed.}
\tablecomments{Parameters for our models which produce the best fit to the \bFW- and \bH-band profiles and SED, for both porous and compact grains.  The grain size distribution is $dN(a)\propto a^\gamma da$, while the surface density is described by $\Sigma(r)\propto r^p$.  Illuminating star: $T_\mathrm{eff}=$3600\,K, $\log g=4.5$ \NextGen\ model, stellar radius 0.88\,$R_\sun$.  Both models give comparable fits to the scattered light and SED, but the measured polarization rules out compact \micron-sized grains (Figs.~\ref{fig:model} \&~\ref{fig:model_JK_pol}).}
\end{deluxetable}

\begin{deluxetable}{rrr}
\tablewidth{0pc}
\tablecaption{Best-Fit Model Avg. Scattering Parameters\label{tab:modelscatparms}}
\tablehead{\colhead{$\lambda$} & \colhead{$g$} & \colhead{$A$}}
\startdata
\sidehead{porous}
606\,nm & 0.83 & 0.52 \\
1.6\,\micron & 0.81 & 0.54 \\
\sidehead{compact}
606\,nm & 0.66 & 0.83 \\
1.6\,\micron & 0.60 & 0.81 \\
\enddata
\tablecomments{Scattering parameters for the outer regions of our best-fit models (Table~\ref{tab:modelparms}), averaged over the grain size distribution.  The scattering asymmetry factor $g$ and albedo $A$ are given.}
\end{deluxetable}

\clearpage

\end{document}